\documentclass[fleqn,usenatbib,useAMS]{rasti}

\usepackage{graphicx}	
\usepackage{amsmath}	
\usepackage{multicol}        
\usepackage{fix-cm}
\usepackage{pdflscape}	
\usepackage[T1]{fontenc}
\usepackage{orcidlink}
\usepackage{makecell}
\usepackage{lastpage}

\newcommand{\mrulefill}{\leavevmode\leaders\hrule height 0.31ex depth -0.3ex\hfill\kern0pt}
\newcommand{\bigO}{\mathcal{O}}

\usepackage[T1]{fontenc}
\usepackage{ae,aecompl}

\usepackage{newtxtext,newtxmath}

\usepackage{bm}

\title[\textsc{speculate}: Emulating and fitting disc wind models]{\textsc{speculate}: A synthetic spectral library, emulator and fitting package for accretion disc winds -- Infrastructure and application to CVs}

\author[A. G. W. Wallis et al.]{
Austen G. W. Wallis$^{\orcidlink{0000-0003-0770-9015}}$,$^{1,2}$\thanks{E-mail: A.Wallis@soton.ac.uk}
Christian Knigge$^{\orcidlink{0000-0002-1116-2553}}$,$^{1}$
James~H.~Matthews$^{\orcidlink{0000-0002-3493-7737}}$,$^{3}$
Knox S. Long$^{\orcidlink{0000-0002-4134-864X}}$,$^{4,5}$
\newauthor
Madeleine-Mai Ward$^{\orcidlink{0009-0009-8478-1396}}$,$^{1}$
Stuart A. Sim$^{\orcidlink{0000-0002-9774-1192}}$,$^{6}$
Adam B. Hill$^{\orcidlink{0000-0003-3470-4834}}$,$^{7,1}$
Amin Mosallanezhad$^{\orcidlink{0000-0002-4601-7073}}$,$^{1}$
Nicolas Scepi$^{\orcidlink{0000-0003-3909-2486}}$,$^{8}$
\\
$^{1}$School of Physics and Astronomy, University of Southampton, Southampton, SO17 1BJ, UK\\
$^{2}$Institute of Astronomy, University of Cambridge, Madingley Road, Cambridge, CB3 0HA, UK\\
$^{3}$Department of Physics, Astrophysics, University of Oxford, Denys Wilkinson Building, Keble Road, Oxford, OX1 3RH, UK\\
$^{4}$Space Telescope Science Institute, 3700 San Martin Drive, Baltimore, MD, 21218, USA\\
$^{5}$Eureka Scientific Inc., 2542 Delmar Avenue, Suite 100, Oakland, CA, 94602-3017, USA\\
$^{6}$School of Mathematics and Physics, Queen's University Belfast, University Road, Belfast, BT7 1NN, UK\\
$^{7}$ComplyAdvantage, 2nd Floor, Fetter Yard, Fetter Lane, London EC4A 1AD, United Kingdom\\
$^{8}$Univ. Grenoble Alpes, CNRS, IPAG, 38000 Grenoble, France
}

\date{Accepted XXX. Received YYY; in original form ZZZ}

\pubyear{2026}

\begin{document}
\label{firstpage}
\pagerange{\pageref{firstpage}--\pageref{LastPage}}
\maketitle

\begin{abstract}

Disc winds are critical components of accreting astrophysical systems and their observations can yield a wide variety of spectral line profiles. Yet, inferring disc wind parameters remains challenging: the underlying parameter spaces are high-dimensional and degenerate, while the radiative-transfer simulations are imperfect and computationally expensive. This paper tackles bridging this gap. Here, we present \textsc{speculate}, an open-source framework for rapidly emulating and fitting observations with synthetic accretion disc wind spectra, generated from the Monte Carlo radiative-transfer code \textsc{sirocco}. \textsc{speculate} includes the release of our cataclysmic variable (CV) training data, a library of $88\,680$ sparsely gridded \textsc{sirocco} spectra spanning the ultraviolet and optical wave bands, a suite of pre-trained emulators, and its interactive \textsc{marimo} user interface. \textsc{speculate} compresses spectral grids using principal component analysis and then provides three approaches to interpolate the resulting latent representation: with a regular-grid interpolation, a neural network, or a Gaussian process. All three architectures enable \textsc{speculate} to generate a \textsc{sirocco} spectrum between $1-40\,\mathrm{ms}$, thereby enabling efficient inference. The first two support rapid $\chi^2$ optimisation. The Gaussian process utilises a Bayesian inference engine coupled with a covariance-aware, flexible likelihood function, adapted and accelerated from the \textsc{starfish} framework. This function helps absorb correlated discrepancies between simulations and data. Tests show all architectures reproduce unseen \textsc{sirocco} spectra with broadly comparable accuracy and can recover their physical parameters. Subsequently, we apply \textsc{speculate} to 5 representative nova-like CV spectra, reproducing several observed line features while also exposing mismatches where the current model grid omits relevant physics.

\end{abstract}




\section{Introduction}

\subsection{Accretion disc winds and their observational signatures}
Outflows from accretion discs are a key component of the physical characterization of many astrophysical sources, from small-scale young stellar objects and cataclysmic variables \citep{Scaringi2026-CVreview} to X-ray binaries \citep{Munoz2026}, tidal disruption events \citep{parkinson_accretion_2020, Gezari2021} and the massive scales of active galactic nuclei \citep{Crenshaw2003, King2015}.  Accretion releases energy, which in turn can launch inflowing mass away from the disc via a variety of physical processes, such as thermal pressure, radiation pressure and/or magnetohydrodynamically driven \citep{Blandford1982, Begelman1983, Proga1998, Proga2000}. The removal of mass and angular momentum can modify the accretion flow (e.g. by limiting the accretion rate), and the outflowing material can also reprocess the radiation produced by the system. Both of these factors can affect the source's observational appearance to a researcher. For example, lower accretion rates can lead to dimmer, cooler discs \citep{shakura_black_1973}. Meanwhile, different ionization states in the wind mean the radiation emitted by the disc can now be absorbed, scattered and re-emitted by the outflow in different proportions \citep{Proga2000}. This variation in the reprocessing of light can lead to significant modifications to the continuum and spectral lines originally emitted from the disc (and central object) that an observer ultimately sees \citep{matthews_impact_2015, Charles2019, Inight2022}.

This ultimately means probing the spectrum of an accreting source is not just observing the flux emitted from an accretion disc and a central source. It also involves examining the complex density, ionization state, geometry, and velocity structure of what could be quite a complex outflow \citep{shlosman_winds_1993, knigge_application_1995, long_modeling_2002}. Given this complexity, interpreting a disc wind-dominant spectrum can be difficult and often requires comparing observations against radiative transfer simulations \citep{knigge_application_1995, long_modeling_2002, higginbottom_line-driven_2014, matthews_impact_2015}.

Spectroscopically, the clearest indication of an outflowing system is blueshifted absorption associated with a spectral line, where material moving towards the observer scatters light away from the observer's line of sight \citep{shlosman_winds_1993, Munoz2026}. These profiles, along with P-Cygni profiles, produce the most direct evidence of outflowing material \citep{Scaringi2026-CVreview}. However, the connection between an observed profile and the wind that produced it is not always this clear-cut. Outflows can also produce a complex variety of other line shapes, ranging from symmetric to asymmetric, and from single- to quadruple-peaked profiles \citep{wallis_diagnostic_2025}. These line strengths and shapes depend on the unique combination of density, ionization state, geometry, velocity fields, and source spectral energy distributions \citep{shlosman_winds_1993, knigge_application_1995, long_modeling_2002, matthews_impact_2015}. Additionally, the appearance of these lines can change with inclination \citep{knigge_application_1995, higginbottom_simple_2013, wallis_diagnostic_2025}. This means multiple distinct outflows may exhibit a nearly identical spectral profile. Meanwhile, the same outflow at a different inclination may result in a substantially different spectrum.

This spectral ambiguity makes it highly unlikely that an outflow's physical parametrization can be diagnosed from a single line profile \citep{Parker2022, wallis_diagnostic_2025}. Typically, constraining an observation's outflow properties requires comparing multiple spectral lines with predictions from a forward model of the outflow \citep{long_modeling_2002, Sim2010, matthews_impact_2015}. Therefore, spectroscopy of a disc wind is an inverse problem in which a researcher attempts to determine which regions of a degenerate, high-dimensional modelling space match the observations -- all while balancing observational uncertainties, model imperfections, and contamination from alternative sources of radiation. 

\subsection{Modelling outflow signatures with \textsc{sirocco}}

To connect the physical properties of an accretion disc wind to observations, we require a forward modelling code. For this, one can use \textsc{sirocco}, a sophisticated 2.5D Sobolev-based Monte Carlo ionization and radiative transfer code that simulates spectra from an azimuthally symmetric accretion outflow \citep{matthews_sirocco_2024}. With \textsc{sirocco}, a researcher can start building connections by systematically searching the parameter space to identify models that reproduce observed line profiles \citep{long_modeling_2002, matthews_impact_2015, matthews_disc_2023, wallis_diagnostic_2025}. Crucially, we have previously used these model grids to analyse observations, develop diagnostics and derive relations that connect physical parameters directly to observed line shapes and strengths. Some examples of these include \citet{matthews_impact_2015, wallis_diagnostic_2025} for cataclysmic variables and \citet{matthews_disc_2023} for quasars. 

However, a critical bottleneck remains: the computational cost of a simulation. A single \textsc{sirocco} simulation takes a minimum of 1--2 CPU-core-hours for an optically thin, diffuse cataclysmic outflow, and up to thousands of CPU-core-hours for optically thick, extremely dense AGN outflows. Ultimately, searching \textsc{sirocco}'s high-dimensional parameter space of 30+ parameters to fit an observed spectrum using traditional grid-search methods is just computationally prohibitive. As a result, studies making use of \textsc{sirocco} tend to limit the scope of parameter exploration to a few key parameters, such as accretion rate, wind mass-loss rate, and wind cone geometry, among others \citep{higginbottom_simple_2013, matthews_impact_2015, matthews_disc_2023, wallis_diagnostic_2025}. Even with this compromise, a researcher will still typically spend months running \textsc{sirocco} numerical simulations to answer their research question; only exploring a tiny portion of the available parameter space. This research strategy would change quite rapidly, however, if a simulation output were near instantaneous. 

\subsection{Emulation}

Emulators, also known as surrogate models, are supervised machine learning models trained to mimic the behaviour of a much more complex and slower simulation. An emulator's output is, at best, only an approximation of the true solution. However, sampling from these models is orders of magnitude faster. They enable us to use slow simulations for direct probabilistic inference. As an example, with \textsc{sirocco}, this is estimating physical properties, such as mass-loss rates and wind geometry, from observed spectra in minutes rather than months.

Emulators are becoming an increasingly common tool across astronomy and cosmology, with some recent examples including, but not limited to radiative transfer -- spanning dust continuum transfer \citep{rino-silvestre_emulart_2022, sethuram_emulating_2023}, accretion disc-wind outflows \citep{matzeu_new_2022}, molecular line transfer \citep{su_emulating_2025}, and planetary atmospheres \citep{tahseen_enhancing_2024} -- stellar and galaxy spectra \citep{czekala_constructing_2015, alsing_speculator_2020, rozanski_transformerpayne_2024, li_cue_2024}, stellar evolution \citep{hon_flow-based_2024}, gravitational-wave backgrounds \citep{laal_deep_2024}, and cosmology \citep{cabayol-garcia_neural_2023, choudhury_gpr-based_2024, saraivanov_attention-based_2024, gunther_uncertainty-aware_2024, gunther_ole_2025, lovick_high-dimensional_2025}. Emulators are commonly built using two main approaches: as neural network architectures or as Gaussian process models. There is no decisive choice between the two; it typically depends on a user's requirements. A neural network tends to perform faster but usually lacks the predictive uncertainty quantification of a Gaussian Process. If both approaches are designed appropriately, their accuracy performance between the two can be similar. Nevertheless, emulation provides researchers with a route to forward-model \textsc{sirocco} simulations for comparison with observations, without requiring them to dedicate vast amounts of work or computational hours.

\subsection{Inference}

Once an emulator is constructed, generating a single evaluation for a given parameter combination is much faster than running the full simulation. This speed makes it practical to find the best-fitting model for a observation via numerical optimization (such as in standard $\chi^2$ minimization). For emulators without intrinsic uncertainty quantification, such as linear interpolation over a grid or an MLP, minimization of this kind is the usual route to a best-fitting point estimate. However, emulators with built-in uncertainty quantification, such as neural network ensembles or Gaussian processes, allow a fuller probabilistic treatment. Their predictive uncertainty can be added to the observational errors in the likelihood, which can then be sampled with \texttt{emcee} \citep{foreman-mackey_emcee_2013} or the No-U-Turn Sampler \citep{hoffman_no-u-turn_2011} to map out the posterior distribution for an observation.

Hence, to accommodate various scientific requirements, our framework implements two distinct inference pathways. Firstly, a more traditional $\chi^2$ optimization method using a lightweight grid interpolator and neural network emulators for immediate inference results. This approach is fast, designed for quick fitting of observations to identify regions of interest in the \textsc{sirocco} parameter space. Secondly, a covariance-aware Bayesian approach utilizing the \textsc{starfish} flexible-likelihood framework for imperfect simulations, coupled with a Gaussian process emulator. This approach is slower, but gives posterior estimates of the \textsc{sirocco} parameters and degeneracies for a given observation that propagate several sources of error. These are: (i) the covariance of measured flux errors. (ii) the Gaussian processes' predictive covariance providing an approximate emulator uncertainty (iii) correlated model-data discrepancies. Any simulation will fundamentally be imperfect, leading to systematic model errors. For example, the simulation's atomic data might be wrong, leading to universally weaker spectral lines for an atomic species, such as carbon. \textsc{starfish} includes a mechanism to account for some of these inconsistencies and thus provides a robuster statistical framework for fitting imperfect models to data.

\subsection{Objectives, scope and contributions}

This paper primarily aims to present the reusable infrastructure of \textsc{speculate}, our emulator-based inference framework for \textsc{sirocco} accretion disc winds. \textsc{speculate} adapts and builds upon the PCA-GP workflow of \textsc{starfish} \citep{czekala_constructing_2015}. We demonstrate how astronomers can use \textsc{speculate} to explore \textsc{sirocco} model grids and fit observed spectra with them, taking nova-like CV observations as a first case study. Previously, exploring the \textsc{sirocco} parameter space, isolating the effect of individual outflow parameters on a spectrum or fitting an observation meant months of expensive, one-off radiative transfer simulations. \textsc{speculate} replaces these simulations with trained spectral emulators, which approximate the output orders of magnitude faster.

The framework supports two inference pathways and three emulator architecture options. Our linear-grid interpolation and neural-network emulators provide lightning-fast $\chi^2$ optimizations and interactive parameter-space exploration. This pathway is also lightweight, so that all astronomers can access these tools, regardless of their computational resources. Our Gaussian process emulators are more demanding and require a GPU. However, these emulators are coupled with the \textsc{starfish} flexible likelihood framework for a posterior inference which accounts for emulator uncertainty and correlated model--data discrepancies. Both paths are supported with a range of interactive tools for inspecting the \textsc{sirocco} synthetic library, training and evaluating emulators, and exploring the spectral effects of different simulation parameters across UV and optical lines.

Our present implementation of \textsc{speculate} is restricted to two grids of high-state (wind-driving) CV configurations, using an axisymmetric KWD wind prescription \citep{knigge_application_1995}, which are most comparable to nova-like CVs. Consequently, CV observations outside of this prescription or with strong contamination from other astrophysical sources that our \textsc{sirocco} physics does not model remains out of scope. These points are touched on in section~\ref{sec:Tier3} and~\ref{sec:discussion}. Nevertheless, we demonstrate fitting some nova-like observations to show how the \textsc{speculate} infrastructure can be applied for any given grid. 

This paper primarily delivers the following: 
\begin{itemize}
    \item The creation of the \textsc{speculate} synthetic library of \textsc{sirocco} CV wind models, consisting of two training grids with two corresponding test grids, totalling $88\,680$ extracted spectra (section~\ref{sec:synthetic library}).
    \item The \textsc{speculate} infrastructure, which includes insights and implementation of a PCA-based spectral representation, an emulator framework with three different architectural approaches, and corresponding inference frameworks for these emulators to use (section~\ref{sec:infrastructure}).
    \item The computational acceleration of the original \textsc{starfish} PCA--GP workflow for higher dimensional model grids (sections~\ref{sec:PCA}, \ref{sec:Emu_GP}, and \ref{sec:accelerate_starfish}).
    \item A rapid lightweight $\chi^2$ optimization inference pathway adjunct to the accelerated \textsc{starfish} Bayesian inference route (section~\ref{sec:inference}).
    \item A suite of benchmark tests to evaluate emulator performance and showcase fitting to observations (section~\ref{sec:Emulator Performance Benchmarks}). 
    \item The release of \textsc{speculate}, including its online availability as a \textsc{hugging face space}, its user guide and its reactive \textsc{marimo} user interface (section~\ref{sec:user guide}).
\end{itemize}
    
\section{Our \textsc{sirocco} CV synthetic library}
\label{sec:synthetic library}

\begin{figure*}
\centering
    \includegraphics[width=\linewidth,height=0.85\textheight,keepaspectratio,
        alt={Eight schematic panels show illustrations of the three sampled values -- low, medium, and high -- of disc accretion rate, wind mass-loss rate, collimation, mass-loss exponent, acceleration length, acceleration exponent, boundary-layer luminosity and boundary-layer vertical extent. The first six parameters define the grid without boundary-layer emission and including the remaining two parameters define the grid with boundary-layer emission.}
    ]{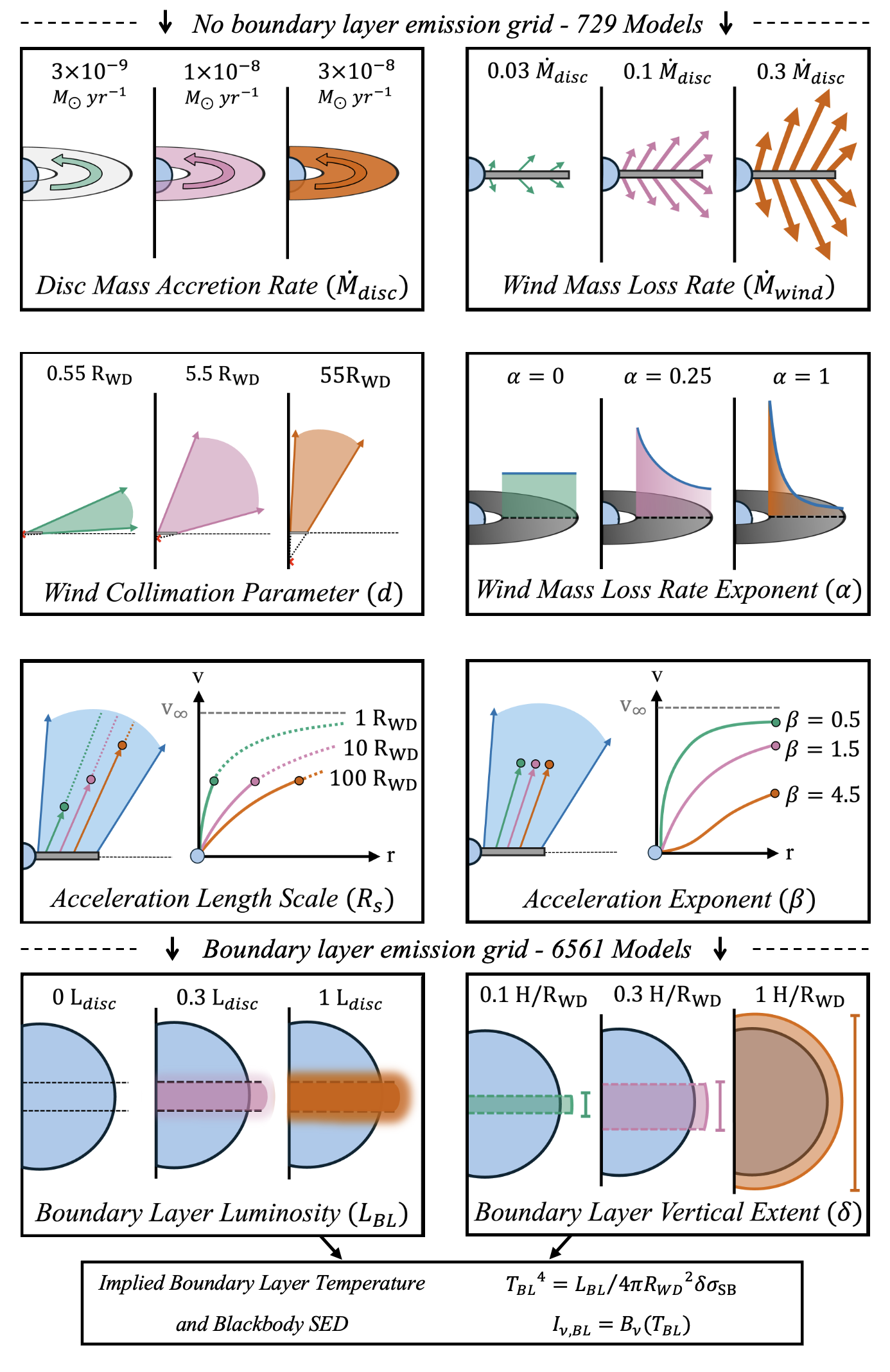}
    \caption{A cartoon schematic of the 8 physical parameters we vary to create our two grids of \textsc{sirocco} cataclysmic variable models. Each panel displays an individual parameter, and each parameter has 3 (labelled) varying values. Each panel has a triadic scheme. The green colour represents a parameter's lower value, the pink colour its middle value, and the orange colour its higher value. The top 6 panels are part of our non-boundary-layer emission CV grid of 729 models. All 8 panels are part of our boundary-layer emission CV grid, which totals 6561 models. NB. Boundary layer vertical extent ($\delta$) is not a \textsc{sirocco} input parameter. However, it \textit{is} an input value for the boundary layer temperature ($T_{\mathrm{BL}}$), which \textit{is} a \textsc{sirocco} input parameter.}
    \label{fig:cartoons}
\end{figure*}

As with training any machine learning model, we rely heavily on the training data. This has led us to create our largest-ever collection of CV outflow models, covering a wide range of physically plausible configurations. We have designed two grids, both of which largely contain similar inputs. However, they differ in that one grid includes boundary-layer emission as a source of radiation, whereas the other does not. The full list of physical inputs for both grids is provided in Table~\ref{tab:combination_table_cv}, with an illustrative schematic in Fig.~\ref{fig:cartoons}. Parameter values typeset in regular font are consistent, fixed values across all models. Parameters in bold font are the inputs which we vary per model. For both grids, we sample 3 values for each varying parameter: a low and a high value on either side of a default (middle) value. These values are designed to create a quasi-regular grid of models. 

\textbf{No boundary layer grid} -- This grid is very similar to the CV grid presented in \citet{wallis_diagnostic_2025}. It still varies the same 6 parameters:
\begin{itemize}
    \item \textbf{Disc mass accretion rate,}
    \item \textbf{Wind mass-loss rate,}
    \item \textbf{The wind's degree of geometric collimation,}
    \item \textbf{The $\alpha$ exponent controlling the wind mass-loss rate,}
    \item \textbf{The acceleration length scale,}
    \item \textbf{The acceleration exponent.}
\end{itemize}
See \citet{wallis_diagnostic_2025} for details. However, we have expanded the number of measured \textbf{inclinations} from 5 to 12, providing finer grading between the different observation angles and creating 5\,103 additional synthetic spectra. These inclinations are set between 30 and 85 degrees in 5-degree increments. Inclination is a particularly important parameter for the output spectrum. The spectral appearance can change quite dramatically depending on whether an observer is looking into the wind cone or from an angle outside the wind cone. We also reduced the number of spectral cycles from 220 to 12, thereby lowering the continuum signal-to-noise ratio of the output spectrum to $\approx50$. This was to accelerate the computational time taken for an individual model, ensuring we could compute the full grid of models within a reasonable time frame. The 6 physical parameters we sample each take on 3 values on our grids, yielding $3^6$ combinations and 729 possible models. With 12 inclinations per model, this leads to a total of $8\,748$ synthetic spectra extracted. 

\begin{table*}
\centering
\small
\setlength{\tabcolsep}{3pt}
\begin{tabular}{l l l l}
\hline
    \textbf{Parameter*} & \textbf{Symbol} & \textbf{Unit} & \textbf{Value} \\
    \hline
    System type & -- & -- & CV \\
    Central object mass & $M_{\mathrm{WD}}$ & $M_{\odot}$ & 0.8 \\
    Central object radius & $R_{\mathrm{WD}}$ & cm & $7.25182 \times 10^8$ \\
    Central object radiation & -- & -- & No \\
    Binary secondary mass & -- & $M_{\odot}$ & 0.4 \\
    Binary period & -- & h & 3.2 \\
    Disc type & -- & -- & Flat \\
    Disc radiation & -- & -- & Yes \\
    Disc radiation type for wind & -- & -- & Blackbody \\
    Disc temperature profile & -- & -- & Standard \\
    \textbf{Disc mass accretion rate} & $\boldsymbol{\dot{M}}_{\mathrm{disc}}$ & $\boldsymbol{M}_{\odot}\,\mathrm{yr}^{-1}$ & $\mathbf{3\times10^{-9}}$, $\mathbf{1 \times 10^{-8}}$, $\mathbf{3 \times 10^{-8}}$ \\
    Disc maximum radius & -- & cm & $2.17555 \times 10^{10}$ \\
    Disc radiation type in final spectrum & -- & -- & Blackbody \\
    \multicolumn{4}{c}{\mrulefill\enspace $\downarrow$ \textit{No boundary layer emission grid -- 729 Models} $\downarrow$ \enspace\mrulefill} \\
    Boundary layer radiation & -- & -- & No \\
    \multicolumn{4}{c}{\mrulefill\enspace $\downarrow$ \textit{Boundary layer emission grid -- 6561 Models} $\downarrow$ \enspace\mrulefill} \\
    Boundary layer radiation & -- & -- & Yes \\
    Boundary layer radiation type for wind & -- & -- & Blackbody \\
    \textbf{Boundary layer luminosity} & $\boldsymbol{L}_{\mathrm{BL}}$ & $\boldsymbol{L}_{\mathrm{disc}}$ & $\mathbf{0,\ 0.3,\ 1}$ \\
    \textbf{Boundary layer temperature (vertical extent)} & $\boldsymbol{T}_{\mathrm{BL}}\,(\boldsymbol{\delta})$ & $\mathrm{K}\,\left(H/R_{\mathrm{WD}}\right)$ & $=f(\boldsymbol{L}_{\mathrm{BL}},\boldsymbol{\delta} = 0.1, 0.3, 1)$ (see equation~\eqref{eq:boundary_layer_temperature}) \\
    Boundary layer radiation type in final spectrum & -- & -- & Blackbody \\
    \multicolumn{4}{c}{\mrulefill} \\
    Wind number of components & -- & -- & 1 \\
    Wind type & -- & -- & Knigge--Woods--Drew \\
    Wind coordinate system & -- & -- & Cylindrical \\
    Wind dimension in x-direction & -- & -- & 30 \\
    Wind dimension in z-direction & -- & -- & 30 \\
    Photons per cycle & -- & -- & $10^7$ \\
    Ionization cycles & -- & -- & 25 \\
    Spectrum cycles & -- & -- & 12 \\
    Wind ionization & -- & -- & Matrix\_pow \\
    Line transfer method & -- & -- & Macro\_atoms\_thermal\_trapping \\
    Atom transition mode & -- & -- & Matrix \\
    Atomic data file & -- & -- & data/h20\_hetop\_standard80.dat \\
    Surface reflection or absorption & -- & -- & Reflect \\
    Wind heating processes & -- & -- & Adiabatic \\
    \textbf{Wind mass-loss rate} & $\boldsymbol{\dot{M}}_{\mathrm{wind}}$ & $\boldsymbol{M}_{\odot}\,\mathrm{yr}^{-1}$ & $\mathbf{0.03,\ 0.1,\ 0.3} \times \boldsymbol{\dot{M}}_{\mathrm{disc}}$ \\
    \textbf{KWD Collimation} & $\boldsymbol{d}$ & $\boldsymbol{R}_{\mathrm{WD}}$ & $\mathbf{0.55,\ 5.5,\ 55}$ \\
    \textbf{KWD Mass loss rate exponent} & $\boldsymbol{\alpha}$ & -- & $\mathbf{0,\ 0.25,\ 1}$ \\
    KWD Velocity factor & $v_{\infty}$ & $v_{\mathrm{esc}}$ & 2 \\
    \textbf{KWD Acceleration length scale} & $\boldsymbol{R}_{\mathrm{s}}$ & cm & $\mathbf{1}$, $\mathbf{10}$, $\boldsymbol{100}\,\times R_{\mathrm{WD}}$ \\
    \textbf{KWD Acceleration exponent} & $\boldsymbol{\beta}$ & -- & $\mathbf{0.5,\ 1.5,\ 4.5}$ \\
    KWD Initial velocity & $v_{0}$ & $v_{\mathrm{c}}$ & 1 \\
    KWD Inner boundary & $r_{\mathrm{min}}$ & $R_{\mathrm{WD}}$ & 1 \\
    KWD Outer boundary & $r_{\mathrm{max}}$ & $R_{\mathrm{WD}}$ & 30 \\
    Wind maximum radius from CV & -- & cm & $10^{12}$ \\
    Wind initial temperature & -- & K & $40\,000$ \\
    Wind filling factor & -- & -- & 1 (smooth) \\
    Number of spectral bins & -- & -- & $14\,400$ \\
    Spectrum wavelength range & -- & Å & 800--8000 \\
    Number of observers & -- & -- & 12 \\
    Spectrum angles / Inclination & $i$ & $^{\circ}$ & 30, 35, 40, 45, 50, 55, 60, 65, 70, 75, 80, 85 \\
    Spectrum orbit phase & -- & -- & 0.5 \\
    Spectrum extraction method & -- & -- & extract \\
    Spectrum type & -- & $\mathrm{erg\,s^{-1}\,cm^{-2}\,\mathring{A}^{-1}}$ & $F(\lambda)$ \\
    Reverb type & -- & -- & None \\
    Photon sampling approach & -- & -- & CV \\
\hline
\multicolumn{4}{l}{*NB: Some parameter labels have been altered to provide shorter, more descriptive names.}\\
\multicolumn{4}{l}{See the documentation for full parameter descriptions and exact labels.}\\
\hline
\end{tabular}
\caption{The default parameters used for the simulation grid. This list follows the same structure as a .pf file typically seen when simulating with \textsc{sirocco}. The regular font parameters are fixed values across all simulations. The bold font parameters are the physical parameter values we vary to create a regular grid of simulations. Each varying parameter has three corresponding grid-point values, yielding $3^{6}$ possibilities (729 total models) for our no-boundary-layer emission grid. For our grid with boundary layer emission, we have $3^{8}$ possibilities (or 6561 models). Note that several parameters, including the likes of the wind mass-loss rate, acceleration length and boundary layer temperature, are dependent upon other input parameters' values. We use a Knigge--Woods--Drew (KWD) wind model for the outflow \citep{knigge_application_1995}. For more detailed information on a particular parameter, see the \textsc{sirocco} \href{https://sirocco-rt.readthedocs.io/en/latest/input.html}{documentation}.}
\label{tab:combination_table_cv}
\end{table*}

\textbf{Boundary layer grid} -- This grid is an expansion upon the previous no boundary layer grid by introducing the CV's boundary layer contribution to the outflow. This addition results in us varying two more key parameters. These are:
\begin{enumerate}
    \item The \textbf{boundary layer luminosity ($L_{\mathrm{BL}}$)} -- This parameter controls the energy radiated away by the boundary layer and its proportional contribution to the final observed spectrum. We adopt three values of $0\times L_{\mathrm{disc}}$, $0.3\times L_{\mathrm{disc}}$ and $1\times L_{\mathrm{disc}}$, which are dependent upon the disc luminosity of a particular model. This disc luminosity, $L_{\mathrm{disc}}$, is calculated following
    \begin{equation}
    L_{\mathrm{disc}} = G  \frac{M_{\mathrm{WD}} \dot{M}_{\mathrm{disc}}}{2R_{\mathrm{WD}}}.  
    \end{equation}
    where $M_{\mathrm{WD}}$ is the mass of the white dwarf in solar mass units, $\dot{M}_{\mathrm{disc}}$ is the disc mass accretion rate in solar mass units per year and $R_{\mathrm{WD}}$ is the white dwarf radius. Naturally, as we increase $L_{\mathrm{BL}}$, the boundary layer will have an increasingly stronger contribution to the total flux of the observed spectra. However, individual spectral lines may strengthen or weaken as the wind's ionization state changes.
    
    \item The \textbf{boundary layer temperature ($T_{\mathrm{BL}}$)} -- The boundary layer in these models is set up to radiate as a blackbody. This parameter controls the effective temperature of that blackbody. However, unlike other gridded parameters, we control this temperature value through another physical parameter that is not among \textsc{sirocco}'s built-in inputs. Instead, we vary the fractional area of the boundary layer covering the white dwarf's surface, also known as the boundary layer thickness ($\delta$). We treat the radial width as a thin shell. We adopt three values of 0.1, 0.3 and 1 for $\delta$ in units of $H/R_{\mathrm{WD}}$, where $H$ is half the full vertical extent of the boundary layer and $R_{\mathrm{WD}}$ is the white dwarf radius. We can connect the boundary layer thickness to the effective temperature through the following equations. The area of a spherical segment is given by: 
    \begin{equation}
        Area = 2\pi R H.
    \end{equation}
    Multiplying by 2 for both hemispheres and introducing the $\delta$ thickness term in units of $H/R_{\mathrm{WD}}$, this equation for boundary layer area becomes: 
    \begin{equation}
    A_{\mathrm{BL}} = 4\pi R_{\mathrm{WD}}^2 \delta.
    \end{equation}
    Boundary layer temperature, area and luminosity are all connected as follows: 
    \begin{equation}
    L_{\mathrm{BL}} = A_{\mathrm{BL}}\sigma_{\mathrm{B}} T_{\mathrm{BL}}^4  
    \end{equation}
    where $\sigma_{\mathrm{B}}$ is the Stefan--Boltzmann constant. Hence, we can write the boundary layer temperature as a function of $\delta$ as follows: 
    \begin{equation}
    T_{\mathrm{BL}} = \left(\frac{L_{\mathrm{BL}}}{4\pi \sigma_{\mathrm{B}}R_{\mathrm{WD}}^2 \delta} \right)^{\frac{1}{4}}.
    \label{eq:boundary_layer_temperature}
    \end{equation}
    Here, $\delta$ sets the effective emitting area used to calculate the boundary layer temperature. However, in \textsc{sirocco}, the boundary layer emission is distributed over the entire white dwarf surface.
\end{enumerate} 

These 2 parameters, together with the no-boundary-layer grid's 6 parameters, each with 3 values, give $3^8=6\,561$ possible models. With 12 inclinations per model ($30^\circ-85^\circ$ every $5^\circ$), this yields $78\,732$ extracted synthetic spectra. However, due to the inclusion of the gridded value $L_{\mathrm{BL}}= 0\times L_{\mathrm{disc}}$, this produces physical overlap with the no boundary layer grid and repeated realizations of the same physical configurations within the boundary layer grid. This choice was deliberately retained to anchor any boundary-layer emulator at the zero-emission limit and to preserve the complete Cartesian design for interpolation across the full boundary-layer thickness axis. This means at $L_{\mathrm{BL}}= 0$, all three $\delta$'s give $T_{\mathrm{BL}}=0$, meaning $\delta$ values are indistinguishable at this boundary. Consequently, $3^6\times3=2\,187$ models (producing $26\,244$ extracted spectra) represent only 729 physically distinct configurations, each with their own separate Monte Carlo realizations. Nevertheless, all realizations are retained, as their distinct coordinates constrain the emulator to recover independence from $\delta$ as $L_{\mathrm{BL}}\rightarrow 0$.

\section{The emulator and inference framework}
\label{sec:infrastructure}

\textsc{speculate} is a framework that allows users to tailor-make their own \textsc{sirocco} emulators to their specific inference needs. \textsc{speculate} adopts three architectures for its emulator design: a linear grid interpolation (GI), a lightweight neural network (NN) surrogate approach, and a Gaussian process (GP) emulator for a probabilistic focused model fitting. The GI and NN methods are designed to enable users to quickly search \textsc{sirocco}'s parameter space for first-pass exploration and can operate with limited computing resources. The GP method, based on the \textsc{starfish} framework, provides an explicit predictive covariance for the emulator's own error in posterior inference. However, for the GP implementation, we do recommend a moderately powerful GPU ($>6\,GB$ VRAM) for high-dimensional (6+ parameters) inferences.  

\subsection{\textsc{starfish}}
\label{sec:starfish}
At its core, \textsc{speculate} was built from \textsc{starfish} \citep{czekala_constructing_2015}, a PCA--Gaussian Process (PCA--GP) emulator-based inference framework. The PCA-GP emulator works as follows: the PCA linear decomposition compresses a spectrum's data points into a handful of latent weights. The GP will then interpolate between these weights so that regions of parameter space without data points have a range of possible weight values corresponding to the GP's confidence bands. The \textsc{starfish} derivation starts from a likelihood comparing reconstructed spectra with the original synthetic grid, introducing $\lambda_\xi$ as the precision of the PCA representation. Here, precision means inverse variance: the PCA representation-error covariance scales as $1/\lambda_\xi$, so larger values imply a smaller assumed uncertainty. Once fitted, a parameter sample can be lossily reconstructed from the GP mean and the PCA basis. A full description of the emulator implementation can be found in the appendix of \citet{czekala_constructing_2015}. 

\textsc{starfish} was a step forward from traditional spectroscopic inference for high signal-to-noise stellar spectra. \textsc{starfish} offered several key improvements. (i) By using Gaussian processes for model interpolation, this meant the parameter posteriors were fundamentally aware of the emulator uncertainty in any particular region of parameter space. (ii) The pixel (residual) covariance was explicitly modelled by Gaussian process kernels instead of assuming independent pixel noise. This is important if there were slight model mismatches between synthetic models and observations. (iii) \textsc{starfish} assumes (and rightly so) that any synthetic model is imperfect. As a result, for any outliers such as large differences in line strength, the code could add a local covariance kernel over these outliers to down-weight the impact of the line mismatches in the MCMC sampling. (iv) The resultant combination of the above leads to a flexible likelihood function that helps to prevent systematic uncertainties in the inference with imperfect synthetic models. 

Crucially, \textsc{starfish}'s methodologies and flexible likelihood function are still at the heart of \textsc{speculate}'s code. However, \textsc{speculate} has developed significantly beyond \textsc{starfish}. Although \textsc{starfish} was built for any synthetic library, \citet{czekala_constructing_2015} uses the \textsc{phoenix} synthetic library \citep{husser_new_2013} of 3 stellar parameters. Our \textsc{sirocco} CV synthetic library consists of up to 9 gridded parameters. With up to $78\,732$ grid points, the computational time required for the original code to operate on our grids was far too great. So instead, \textsc{speculate} now has a GPU-accelerated code version for these high-dimensional outflow inferences, an expanded range of possible emulator architectures, and a plethora of further diagnostic tools for \textsc{sirocco} spectra. Additionally, we have designed \textsc{speculate} with a reactive marimo notebook \citep{agrawal_marimo_2023} user interface for easier navigation and use. Subsequently, the following sections will outline changes from the original \textsc{starfish} implementation, important points for improving emulator performance and a short user guide. 

\subsection{Decomposition methods}
\label{sec:decomposition methods}

A single \textsc{sirocco} spectrum in our synthetic library spans $\sim10^{4}$ wavelength bins. With $\sim10^{4}$ spectra in the whole library, a direct GP interpolation in flux space is far beyond current computational hardware capacity. GP cost scales with the number of features (wavelength bins), so we need to shorten these flux vectors (spectra) into a significantly smaller vector space that a GP \textit{can} actually model. 

To do this, we turn to matrix decomposition methods. These are a collection of techniques designed to break down a complex matrix into simpler parts, which, depending on the need, can make computing tasks much more efficient. For our needs, this is dimensionality reduction: reducing the flux matrix size for our GP in exchange for some information loss from the input matrix. The Python package \textsc{scikit-learn} \citep{pedregosa_scikit-learn_2011} provides a number of methods for dimensionality reduction, including the commonly used principal component analysis (PCA), independent component analysis (ICA), and non-negative matrix factorization (NMF). There are other machine learning methods for dimensionality reduction, such as autoencoders, but for the moment, we restrict ourselves to linear decomposition methods. \textsc{starfish} shipped with PCA at the heart of the code, and \textsc{speculate} continues to use PCA for two key reasons, which will be briefly outlined here and in section~\ref{sec:PCA}. 

Firstly, using a decomposition method ultimately means our emulator's performance will always have an error floor due to information lost by the exact method in question. No matter how well we can predict the weight of the decomposition, we are ultimately limited by the method's reconstruction accuracy in turning eigenspectra and their corresponding weight coefficients back into flux space. Therefore, the better the reconstruction performance of the decomposition method is at decomposing and rebuilding our synthetic library, the better our emulator performance limit can be. Hence, we tested four different `drop-in' methods to compare to PCA on a theoretical grid of spectral line profiles. These line profiles included P-Cygni, absorption and emission lines. The different methods were SparsePCA, Truncated Singular Value Decomposition, Independent Component Analysis and Non-negative Matrix Factorization. In practical terms, there was no meaningful improvement in RMSE scores between the reconstructed spectra and their original counterparts when using one of these alternative methods compared with PCA. Reporting the exact details of this study is therefore out of scope here. However, secondly, the deciding factor in keeping PCA was its orthogonal basis.

\subsection{Principal component analysis}
\label{sec:PCA}

\subsubsection{The PCA representation}
\label{sec:The PCA representation }
PCA is a dimensionality reduction method that simplifies higher-dimensional data while retaining as much variance as possible for a given number of components. The idea behind PCA is to project the data onto a new orthogonal basis that best describes its variance. \textsc{scikit-learn} implements PCA through a singular value decomposition (SVD), which indirectly breaks the dataset matrix (our synthetic library) into a collection of eigenvectors, coefficients and eigenvalues. Following terminology used in \citet{czekala_constructing_2015}, these are the `eigenspectra', `weights' and `variance-explained by each component'. Thus, we can represent an individual \textsc{sirocco} spectrum as a particular linear combination of an eigenspectrum and weights. These eigenspectra are ordered hierarchically by decreasing variance-explained. As a result, truncating them to the first $M$ principal components, where most of the data's variance usually exists, reduces the required data to represent \textit{most} of the original input data. 

Mathematically, this process is as follows. We arrange our spectral library into a flux matrix of shape $\mathbf{F} \in\mathbb{R}^{N \times D}$, where $N$ is the number of grid points (number of spectra), and $D$ is the number of wavelength bins. After centering and standardizing our library to the matrix $\mathbf{Z} \in \mathbb{R}^{N \times D}$, its SVD matrices are given by
\begin{equation}
    \mathbf{Z}=\mathbf{U}\,\mathbf{S}\,\mathbf{V}^{\top}.
\end{equation}
Truncating to the first $M$ principal components, where $M \leq \min(N-1,D)$ and typically $M \ll D$, our $\mathbf{Z}$ matrix can now be defined by our weight matrix $\mathbf{W}_M \in \mathbb{R}^{N \times M}$ and eigenspectra matrix $\boldsymbol{\Xi}_M \in \mathbb{R}^{M \times D}$, where each is given by
\begin{equation}
    \begin{aligned}
        \mathbf{Z} &\approx \mathbf{Z}_M \equiv \mathbf{W}_M\,\boldsymbol{\Xi}_M,\\
        \mathbf{W}_M &= \mathbf{U}_M\,\mathbf{S}_M = \mathbf{Z}\,\mathbf{V}_M,\\
        \boldsymbol{\Xi}_M &= {\mathbf{V}_M}^{\top}.
    \end{aligned}
\end{equation}
This means we can reconstruct any spectrum by the linear combination of its weight ($w_{i\ell}$) and eigenspectrum ($\xi_{\ell j}$), following
\begin{equation}
    z_{ij} \approx \sum_{\ell=1}^M w_{i\ell}\, \xi_{\ell j}
\end{equation}
where $i$, $j$, and $\ell$ are the row, column, and principal component index, respectively. 

Using PCA is fantastic, as we can now reduce the data size of our synthetic library while retaining most of the variance in the preprocessed training library. Using an A100 GPU to compute the PCA is typically on the order of a second for our synthetic library. A CPU fallback is of a similar, but slower, order\footnote{The PCA decomposition computational time processing the CV boundary layer grid on an A100 GPU takes $1.38\,\mathrm{s}$, and $3.40\,\mathrm{s}$ on an Intel Xeon Gold 6338 CPU.}. Hence, we can effectively deploy the algorithm in \textsc{speculate}.

\subsubsection{The computational scaling of the PCA--GP}
\label{sec:The computational scaling of the PCA--GP}

Now, we progress to the major benefit of keeping PCA; its orthogonal basis can help reduce the peak memory load of the GP calculation. In scaling \textsc{starfish} towards higher-dimensional inference, the dense $\mathbf{V}_{11}$ covariance matrix described in \citet{czekala_constructing_2015} would have dimensions of $[NM\times NM]$. The size of $\mathbf{V}_{11}$ is the limiting factor in scaling the GP. This meant the more input data we wanted to include, and the more accurate we wanted the emulator to be, the GP's computational complexity ($\bigO((MN)^3)$ in time and $\bigO((MN)^2)$ in memory) grew quickly infeasible. However, the orthogonality of the principal component basis means this $\mathbf{V}_{11}$ is block diagonal and can be separated into $M$ covariance blocks as 
\begin{equation}
    \mathbf{V}_{11} = \mathrm{blockdiag}(\mathbf{V}_{11}^{(1)}, ..., \mathbf{V}_{11}^{(M)}),
\end{equation}
where each block $\mathbf{V}_{11}^{(\ell)} \in \mathbb{R}^{N\times N}$ is the GP covariance matrix for a single PCA component $\ell$. This means for a given component's weight vectors $\mathbf{w}_\ell$, the log-likelihood can be computed as a sum of the form
\begin{equation}
    \ln\; \mathcal{L}= -\frac{1}{2}\sum_{\ell=1}^M \left[\mathbf{w}_{\ell}^{\top}\left(\mathbf{V}_{11}^{(\ell)}\right)^{-1}\mathbf{w_\ell} + \mathrm{ln\,det}\,\mathbf{V}_{11}^{(\ell)}\right],
\end{equation}
dropping constant terms. Following the upstream \textsc{starfish} training implementation, we optimize this projected-weight likelihood without the adjusted $\lambda_\xi$ prior described in \citet{czekala_constructing_2015}. This means \textsc{speculate} can construct and process blocks sequentially, removing our biggest bottleneck and reducing our peak memory usage to $\bigO(N^2)$. In time, each block still requires an exact dense Cholesky factorization and scales linearly with the number of principal components as $\bigO(MN^3)$. This means care should still be taken not to unnecessarily max out the number of principal components in \textsc{speculate}. A full investigation into the estimated number of PCA components a user should use is presented in section~\ref{sec:choosing_pca_components}. However, the maximum is reasonable and typically on the order of a few hours for entry-level GPUs using our no-boundary-layer CV grid. For the larger boundary-layer CV grid, users will still struggle with GPU memory capacity, as $N$ is an order of magnitude larger. 

\subsubsection{Spectral preprocessing}
\label{sec:PCA Spectral Preprocessing}

\textsc{speculate} offers several preprocessing options for the input \textsc{sirocco} spectra prior to the PCA decomposition. This includes the following. (i) The flux space in which the trained emulator wants to operate. The options are `\texttt{linear}' to train the emulator in sirocco's flux space, erg\,s$^{-1}$\,cm$^{-2}$\,Å$^{-1}$ units at 100\,pc; `\texttt{log}' to train the emulator to operate in log space, or `\texttt{continuum normalized}' space which is helpful when the user is unaware of the distance to the source object. A choice here should depend on which flux space a user's observation is in and which line shapes are present. For example, many sharp spectral features may benefit from preprocessing in log space. (ii) The option to smooth the input (training) \textsc{sirocco} spectra. This option can smooth the input sirocco spectra using a 1-D Gaussian filter with a $\sigma=2$ kernel, thereby decreasing the Monte Carlo noise. The trade-off is tiny modifications (broadening/weakening) to the spectral shapes and a reduction in strength for very sharp spectral features. However, removing photon noise from the spectra means, in principle, the PCA should not `waste effort' in attempting to fit this excess noise. We always tend to use this option. (iii) The wavelength range of the \textsc{sirocco} spectra. As will be discussed next in section~\ref{sec:reconstruct_performance}, the number of spectral lines included in the emulator input data has a direct impact on the number of PCA components that should be used. Care should be taken not to attempt to fit the full UV range to the optical range at once, as an extreme number of PCA components and samples would be required. Equally, not providing too many PCA components to a single spectral line emulator, where excess components will begin attempting to include spectral noise. 

Once these choices are made, each spectrum is mean-normalized, and the resulting synthetic library is centred and scaled to unit variance following standard PCA preprocessing procedures. For reconstruction, the reverse transformations are applied to give the displayed emulated spectrum.

\subsection{PCA reconstructive performance on \textsc{sirocco} spectra}
\label{sec:reconstruct_performance}

To achieve an exact reconstruction of all \textsc{sirocco} spectra in the $\mathbf{Z} \in \mathbb{R}^{N \times D}$ matrix, we would be required to keep all nonzero principal components. However, for \textsc{speculate} to be practical, we require a significantly smaller number of components than this. Truncating the PCA basis to a certain number of components gives a lossy approximation. While \textsc{speculate} can report the variance-explained score before training, and a user may be able to maintain a high percentage of the total variance, any truncation will inevitably introduce a reconstruction error floor that propagates into every downstream inference. There is also no guarantee that a high variance-explained score translates into an accurate reconstruction. Therefore, a user must be aware of how this lossy approximation manifests itself in a reconstructed spectrum. Also, here we provide a rule-of-thumb estimate of how many PCA components one should expect to use for a given spectrum.

\subsubsection{PCA truncation artefacts}
\label{sec:pca_truncation_artefacts}

\begin{figure*}
\centering

	\includegraphics[width=\linewidth,height=0.74\textheight,keepaspectratio,
        alt={A toy spectrum that contains an absent line, an absorption feature, an emission feature, a P-Cygni profile and a double peak. PCA reconstructed spectra using progressively more components are shown to more accurately reconstruct the toy spectrum; the moving sharp blue edge of the P-Cygni profile is difficult to reproduce with few components.}
    ]{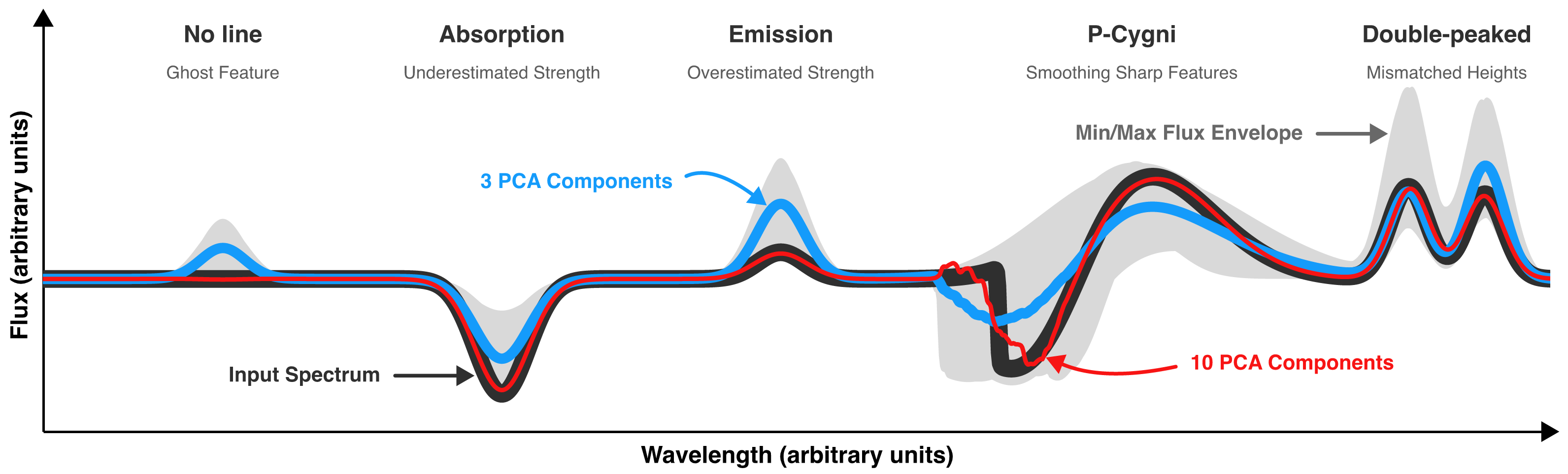}
    \caption{A schematic illustration of how PCA \textit{can} distort different common spectral line shapes. The comparison shows one spectrum (as a thick black line) from a grid of 125 toy spectra (shown by the grey min/max flux envelope) against two PCA reconstructions of that input spectrum, truncated to the first 3 principal components (as a moderately thick blue line) and to 10 principal components (as a thinner red line).}
    \label{fig:line_shapes_pca}
\end{figure*}

To test the reconstruction performance of PCA for spectra, we employ toy grids of spectral line shapes, pass them through the PCA process, and then back. We then measure the RMSE between the reconstructed spectra and their original counterparts and assess the scores across varying numbers of PCA components. We can use this method to discover areas of weakness and information loss with PCA. Fig.~\ref{fig:line_shapes_pca} is an example of one of these toy spectra grids passed through PCA. We have a sample spectrum from the toy spectral grid, shown as a thick black line, comprising a collection of commonly seen line-profile shapes from CV systems that may indicate outflowing properties. These include absorption lines, emission lines, P-Cygni profiles, and double-peaked lines. Surrounding the spectrum is a grey band, which shows the minimum and maximum ranges of the other 124 spectra in the grid. This collection of spectra is then passed through the PCA calculation. We reconstruct the plotted spectrum, truncated to 3 PCA components (blue, moderate thickness) and to 10 PCA components (thinnest red line). The 3-PCA component line represents selecting too few components for 5 spectral lines, and the 10-component line shows the impact on the reconstructed spectrum by increasing the number of components. While our tests analyse all sample spectra across a range of possible combinations, the spectrum shown here is designed to highlight typical weaknesses in PCA's reconstruction for each spectral line type. 

\textbf{The Ghost Feature} -- Let us start with the `no line' feature on the left-hand side. This scenario depicts an absent line at the wavelength position of a common line one would typically expect to see. For example, this could be H\,$\beta$ or a \ion{He}{ii} emission line. In a denser wind model, we may see strong emission in these lines. However, with a lower wind density or different ionization state, this emission may be weak or absent \citep{matthews_impact_2015}. We include this behaviour in our toy grid, but in this particular sample, there is no line. The impact is that if there are too few principal components, the emulator can produce a reconstruction artefact, a non-existent line, showing a `ghost' profile from the other very strong lines seen in the input grid. Increasing the number of components can reduce this effect. 

\textbf{Over/Underestimated Strength} -- Next are the `absorption and emission line' features. These scenarios depict lines that are typically always present on a spectrum in some form or another. However, in the absorption line, the sample spectrum is a strong absorption feature that is unusual in our grid space. For the emission line, the line is weaker than we typically see in our grid space. If we have too few PCA components, the emulator can tend to underestimate or overestimate the line strength, effectively preferring an average line strength over the input grid. Again, increasing the number of components can reduce this effect. 

\textbf{Uneven Double Peaks} -- Penultimately, on the far right of the spectrum, we have `double-peaked lines'. These can be either both emission or both absorption, although the two features may have different amplitudes. This situation can manifest through two mechanisms: disc rotation splits an emission line from its rest wavelength, due to seeing blue and redshifted velocities at high inclinations. Alternatively, this shape can arise from two spectral lines whose rest wavelengths are very close to each other, for example, the \ion{Si}{ii} doublet. Again, as with absorption and emission lines, there is a tendency for PCA with a small number of components to take the average of the input grid, leading to under- or overestimation depending on whether the peak is stronger/weaker than the average input grid amplitude. However, increasing the number of components can reduce this effect.

\textbf{Smoothing Over Sharp Edges} -- Then, finally, we have the `smoking gun' of outflowing systems, a `P-Cygni' profile. This line profile's blueshifted absorption is an extremely valuable indicator of material being ejected from the CV toward an observer. Photons emitted isotropically get intercepted by this material, creating an absorption line feature exclusively on the blueshifted side of the emission line. The wavelength position of the blue edge indicates the fastest projected velocity of the material responsible for the observed absorption. This position also depends on the wind's ionization state. A sharp edge can occur when the line remains optically thick up to the projected terminal velocity. This sharp edge is particularly difficult for PCA to reconstruct, as changes in projected velocity can cause the edge to translate across wavelength space. In PCA terms, this means we have our edge translating across the input dimension feature space ($\lambda_N$ in our flux matrix). This means that, for us to represent all spectra across our grid of spectra accurately, we require a significantly higher number of components to represent every feature column of the matrix in which one of these edges appears. A stark example would be if every spectrum in our input grid had a P-Cygni blue edge at a different wavelength bin. We would require that all nonzero principal components accurately represent this line shape. This makes it difficult for PCA to reconstruct these shifting edges with only a small number of components. For spectral lines that remain close to their rest wavelengths across a grid, this is less of a problem. However, in our \textsc{sirocco} outflow models, changes in the projected wind velocity can often move the blue absorption edge across wavelength space.

So how does one approximate a grid of P-Cygni profiles? A P-Cygni profile can be broken down into two parts: the red emission/blue absorption Gaussian-like profiles, and the sharp blueshifted edge. The Gaussian-like part of the profile behaves similarly to a regular emission/absorption line. However, the blueshifted (discontinuous) edge is `averaged' into a smooth, continuous shape in low-dimensional principal component representations. When the number of principal components is increased, the gradient of the line over the discontinuous edge steepens, and the feature is present over a smaller wavelength range. However, this part of the reconstructed spectrum behaves more like a step function. The easiest comparable case is passing a `square line' through PCA; the result is very reminiscent of the ringing seen in the Gibbs phenomenon in signal processing. The higher the number of PCA components, the closer the reconstructed spectra look `square', but you additionally get increasing oscillations the closer you get to the transition point of the square shape. 

\subsubsection{Choosing the number of PCA components}
\label{sec:choosing_pca_components}

Finally, the question now becomes: `How do we mitigate the impact of this poor representation?' To answer this, we generated numerous toy spectral grids, passed them through PCA and back, and then measured how many principal components we needed to achieve a 95 per cent variance-explained score and a reconstructed spectrum Normalized RMSE error of less than 2 per cent. These toy grids, through PCA, consisted of pure square, triangular, Gaussian, and P-Cygni spectral lines. We also test composite spectra with mixed full-width half-maximum Gaussians and a composite spectrum like the spectrum shown in Fig.~\ref{fig:line_shapes_pca} with added Gaussian noise. The latter composite spectral grid is most representative of a true observation. Each grid contained 50 spectra with 655 wavelength bins, ranging from including only one line to up to ten lines per spectrum. For each grid, the number of principal components tested increased from one per spectral line to ten per line, capped at 49 components in total. The results from this experiment can be seen in Table~\ref{tab:PCA_components_needed}. These tests suggest that smooth spectral lines need relatively few PCA components, while sharp edges and more complex profiles generally require more.

\begin{table}
\centering
\small
\setlength{\tabcolsep}{3pt}
\begin{tabular}{lccc}
\hline
        \textbf{Line Profile Shape} & \textbf{Mean} & \textbf{Range} & \makecell{\textbf{Recommended}\\\textbf{multiple per line}}\\
        \hline
        Gaussian (Smooth Line)& $1.1\times$ & $1\text{--}2\times$ & $2\times$\\
        Mixed Width Gaussians & $1.1\times$ & $1\text{--}2\times$ & $2\times$\\
        Triangular (Sharp Line) & $1.8\times$ & $1\text{--}3\times$ & $3\times$\\
        P-Cygni & $2.1\times$ & $1\text{--}4\times$ & $4\times$\\
        Mixed Profiles & $2.6\times$ & $1\text{--}7\times$ & $3\times$\\
        Square & $5.3\times$ & $3\text{--}9\times$ & $9\times$\\
\hline
\end{tabular}
\caption{An initial rule-of-thumb multiplier for the number of principal components needed to represent spectral lines of different generalized shapes. The mean and range are the average and minimum--maximum multipliers required to achieve a 95 per cent variance-explained score and a Normalized RMSE error that is less than 2 per cent across the toy grids of a particular line shape.}
\label{tab:PCA_components_needed}
\end{table}

\subsection{Interpolating the PCA weight space}
\label{sec:emulators}

\begin{figure*}
\centering
    \includegraphics[width=\linewidth,height=0.74\textheight,keepaspectratio,
        alt={Three slices through PCA weight space compare linear grid interpolation, neural-network interpolation and Gaussian-process interpolation against the input weights. The slices vary wind mass-loss rate, acceleration length and inclination. Shading shows Gaussian-process uncertainty.}
    ]{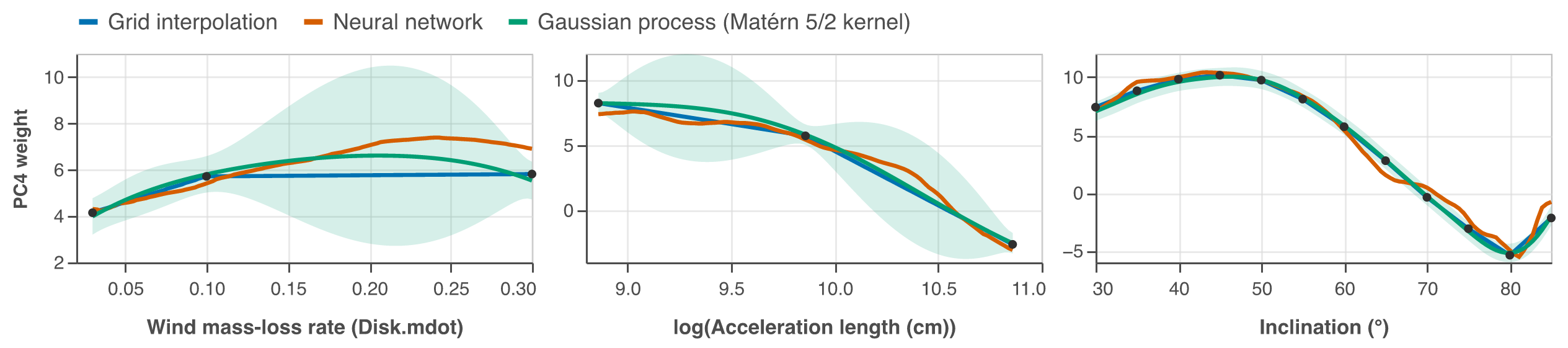}
    \caption{Interpolation examples of the PCA weights for principal component 4 using \textsc{speculate}'s 3 different emulator methods: the grid interpolation method in blue, the neural network method in orange, and the Gaussian process with a Matérn-5/2 kernel in green with its faded green $2\,\sigma$ confidence band. For each 1-D slice, the specified parameter, wind mass-loss rate, acceleration length or inclination, varies while every other parameter in the grid is fixed at its middle gridded value. The black data points are the original grid weights. Further inspection of this weight space can be achieved using \textsc{speculate}'s diagnostic tools in the \texttt{quick fit} (for GIs and NNs) and \texttt{training tool} (for GPs) notebooks. This particular plot's emulator configuration utilized a wavelength range of $1150\text{--}1750\,\mathring{A}$ on smoothed data in their original flux space from our full no-boundary-layer emission CV grid.}
    \label{fig:interpolation}
\end{figure*}

Fundamentally, emulators can be thought of simply as \textit{fancy} interpolators. We could interpolate between grid points directly in flux space. However, this will typically be very computationally intensive, the opposite of what an emulator is designed to be. Instead, we use our PCA decomposition method, which allows us to map the weight space more efficiently. Now, for a new parameter vector (combination), we predict the corresponding specific weight vector $\mathbf{w}_i$, combine with the fixed eigenspectra, apply the inverse pre-processing steps, and retrieve a brand new spectral prediction for that new parameter combination. Ultimately, \textsc{speculate} provides three different methods for this interpolation: a regular grid interpolation, a neural network approach, or a Gaussian process, each of which is shown in Fig.~\ref{fig:interpolation} on a small slice through the weight space. In \textsc{speculate}, users can view these emulator interpolations through the weight space more thoroughly after training. However, there are two important points to note when looking at these plots. (i) Each method imposes different assumptions; in particular, the NN and GP are not constrained to reproduce every training weight exactly. (ii) Principal components are ordered by variance-explained. As a result, the weights can also span different dynamic ranges. Therefore, visually, larger principal components with larger absolute weight values can \textit{look} to fit better than other components with smaller dynamic ranges. However, the error difference between them should be fairly close to each other in absolute terms.

\subsubsection{Emulating with a regular-grid linear interpolation}
\label{sec:Emu_GI}
A regular-grid linear interpolation of the weight space is the most basic approach. We assume that the connecting parameter space between two weights is linear, which is extremely unlikely, particularly given our other methods. So why would we use such a method when we likely know that interpolating this weight space is not linear? The answer is speed. Creating a GI emulator is near-instantaneous, and inference can sometimes be completed in under a minute. To sample (emulate) a spectrum, we take weight values directly from this linear line for the PCA reconstruction. This is shown as the blue line in Fig.~\ref{fig:interpolation}. Emulators, by their nature, are approximate, so making this assumption for speed is a fair trade-off. With the GI method, our reconstruction error is the information lost due to PCA and the linear weight-space assumption. The overall RMSE impact of GI on the reconstructed spectra can also be seen in the blue line in Fig.~\ref{fig:train-RMSE-Tier1} and ~\ref{fig:Test-RMSE-Tier1}, and compared to the PCA truncation error in black. 

\subsubsection{Emulating with a neural network}
\label{sec:Emu_NN}
Neural network interpolation is the first of our two methods and assumes that the interpolation is not necessarily linear. However, we must maintain the assumption that the interpolation is continuous. Our neural network implementation can operate in three ways: a single neural network with a fixed architecture, an ensemble of 3 neural networks, or an \textsc{optuna} \citep{ozaki_optunahub_2025} study of neural network architectures.

Firstly, our single neural network option is the simplest of the three. It is designed to be a lightweight (fewer than $250\,000$ model parameters) feed-forward neural network for faster training and inference. This option has a fixed architecture of

\begin{itemize}
    \item Number of neural layers -- 4
    \item Width of a neural layer -- 256 neurons
    \item Width pattern -- Constant size
    \item Activation function -- ReLU
    \item Dropout Rate -- 1 per cent
    \item Dropout Strategy -- Increasing
    \item Batch Normalization -- True
    \item Batch Size -- 128
    \item Skip Connections -- False
    \item Optimizer -- AdamW
    \item Learning Rate -- $1.5\times10^{-3}$
    \item Learning Rate Scheduler -- Reduce on plateau
    \item Weight Decay -- $5\times10^{-3}$
    \item Output Normalization -- Standard
\end{itemize}

determined through 4 rounds of a 30-trial \textsc{optuna} study on a particular emulator, narrowing the scope of the investigation with increasing round number. During hyperparameter selection, the grid was randomly split 90:10 for training/validation subsets after the pre-processing/PCA steps, and the architectures were ranked according to their validation mean-squared error in standardized PCA-weight space. This is so we can compare how well different neural networks predict weights on that fixed basis. The trained emulator was then evaluated on the held-out test grid. Once complete, this final architecture was selected for its balance of accuracy and low number ($209\,000$) of trainable parameters. Subsequently, when choosing the single neural network option and retraining with this architecture, \textsc{speculate} will use the entire \textsc{sirocco} grid as training data. This is because the grid is sparse, and splitting it will unnecessarily harm parameter-space coverage for a trained emulator. Nevertheless, the generalized performance of the trained emulator is evaluated post-training using the corresponding \textsc{sirocco} test grid, which should highlight areas of weakness for researchers. Ultimately, this may not be the \textit{best} architecture for \textit{all} emulator configurations. However, this architecture maintained strong performance across configurations during \textsc{speculate} development. Additionally, comparing the neural network's RMSE (shown as an orange line) with the PCA's truncation floor error in Figs.~\ref {fig:train-RMSE-Tier1} and~\ref{fig:Test-RMSE-Tier1} shows that further optimizing the architecture is likely to yield diminishing returns. Fundamentally, we are limited in our data availability instead.

Our next option is a neural network ensemble. This setup runs 3 neural networks with the same architecture as the single neural network option. However, the difference lies in how we pass data through the network during training. Instead of using the full \textsc{sirocco} grid, we split the data into a 90:10 training--validation set. Each member of the ensemble uses a different training--validation split and random seed. This means that differences in best fits across neural networks indicate an `uncertainty' in the parameter estimates. The uncertainty here is quantifying more of the data/emulator uncertainty rather than systematics in the \textsc{sirocco} models. It gives an idea of the \textit{sensitivity} of the parameter results. In the future, this ensemble option may expand to include alternative neural architectures, for example, changing activation functions/number of layers/width per layer for members. 

Our final option is the \textsc{optuna} search to tune the neural network's hyperparameters. A user may find a region of flux-wavelength space where the single neural network model severely underperforms. In this case, this option is a mechanism for a user to find a better architecture. An \textsc{optuna} trial requires a training--validation data split, so it follows a similar protocol to the ensemble. Underperforming models are pruned from the trial list until an optimal setup is discovered. The longer the trial list, the longer the search. However, more of the later-stage models will be pruned based on the results of earlier trials, so increasing the number of trials is likely not to increase the overall computational time linearly. 

\subsubsection{Emulating with a Gaussian process}
\label{sec:Emu_GP}

Our Gaussian process mapping of input parameters to weights closely follows the \textsc{starfish} implementation as described in section~\ref{sec:starfish}. We have, however, made three important options available to help \textsc{speculate} scale to higher dimensions. Firstly, \textbf{the optimizer}. \textsc{starfish} was built with \textsc{scipy}'s minimizer for one or more variables using the Nelder-Mead algorithm \citep{nelder_simplex_1965, gao_implementing_2012}. This method remains a part of \textsc{speculate}. However, there are now two additional optimizers included; the L-BFGS-B quasi-Newton method \citep{byrd_limited_1995, zhu_algorithm_1997} and the CMA-ES derivative-free evolutionary strategy (see the tutorial by \citet{hansen_cma_2016}).
While Nelder-Mead is considered a robust optimizer for finding local minima, in higher dimensions, even with the adaptive Nelder-Mead simplex of \citet{gao_implementing_2012}, the method can become ineffective. Their study does not establish a fixed number of dimensions beyond which Nelder-Mead becomes ineffective. However, their testing capped out at around 60--100 dimensions when reaching this conclusion. 

So how many optimizable dimensions do we encounter in \textsc{speculate}? Well, as \textsc{starfish} optimized all GPs simultaneously with a $\lambda_{\xi}$ noise term (precision of the eigenspectra basis representation), if we have $P$ parameters and $N$ principal components, the number of hyperparameters passed to the optimizer scales as $P\times N + N + 1$. This equation is from $P\times N$ hyperparameters describing the GP length scales, $N$ hyperparameters describing the variance for each principal component, and the additional $\lambda_{\xi}$ term for the $+1$. So, for example, our no-boundary-layer emission CV grid with 7 parameters and 20 principal components will yield 161 hyperparameters. Meanwhile, scaling to up 30 components, the number used for our released pre-trained emulators, contains up to 241 hyperparameters. Ultimately, in this 100+-dimensional regime, we found that Nelder-Mead frequently failed.

Our first step in solving this issue was with the CMA-ES optimizer. While this method will still encounter problems at higher dimensions, it is hardware and time limitations that ultimately cap the method's effectiveness, with each generated search point scaling $\bigO(n^2)$ in memory and time \citep{hansen_cma_2016}. In \citet{loshchilov_lm-cma_2015}, they considered this limit to be of $n>1000$ dimensions (before subsequently creating a low memory version of CMA-ES to replace L-BFGS). During development, we found that CMA-ES was more effective as an optimizer during training for matching configurations. It is also important to note that the dynamic ranges of all parameters should be scaled to be near unity to prevent both PCA and the optimizers from being dominated by a high-dynamic-range term. Subsequently, \textsc{speculate} moved to utilize CMA-ES as its default optimizer. As CMA-ES and Nelder-Mead both take significant time to optimize, L-BFGS-B is also included as a faster alternative. However, it should be noted that there is a higher risk of L-BFGS-B falling into local minima in a fairly complex likelihood landscape. For training \textsc{speculate} emulators, we recommend restricting this method to cases where a small number of hyperparameters are being optimized. For example, $n<10\text{--}20$. 

However, even with a `better' optimizer, trusting any optimizer to find the true global minima on a $100+$ dimensional log-likelihood landscape is a big ask. This brings us to our second option: \textbf{per-component training} where we optimize each principal component's hyperparameters individually, with a bounded Brent minimization \citep{brent_algorithms_1973} in $\ln(\lambda_{\xi})$ afterwards to perform a derivative-free search for a $\lambda_{\xi}$ value (the shared hyperparameter of the principal components) that minimizes the total negative log-likelihood. This has the advantage that the optimizer will only minimize $P + 1$ hyperparameters at any one time, i.e. no more than $10$ hyperparameters for \textsc{speculate}'s current largest grids. This is a much more achievable minimization problem for any optimizer. From this point, a user can use these trained emulators. The block-diagonalization of $V_{11}$ means the log marginal likelihood decomposes into component-wise terms while $\lambda_\xi$ is held unchanged during per-component training. However, the subsequent Brent refinement of $\lambda_\xi$ can shift the component optima, so this sequence is not necessarily equivalent to joint optimization. A user can therefore use the \textbf{continue training} option with per-component training and refine $\lambda_\xi$ options disabled to jointly optimize all component hyperparameters and $\lambda_\xi$ from this initial solution, potentially making subsequent joint optimization easier.

Thirdly, we have allowed Matérn-3/2 and Matérn-5/2 kernels in addition to \textsc{starfish}'s original radial basis function (rbf) kernel (also known alternatively as square exponential kernel). \textsc{starfish} made the deliberate choice to use the RBF kernel because of the smoothness assumption of the weight space. However, for disc wind spectra, this smoothness is not as reliable an assumption to make. Inspecting neural network weight fits for principal components shows significant variability. Additionally, inspecting the more densely sampled inclination parameter in Fig.~\ref{fig:interpolation} shows that the gradient in weight space can change significantly over a small section of parameter space. This is reasonable with a denser sampling. However, if we only use our sparse sampling of the inclination option at $35^{\circ}$, $55^{\circ}$, and $80^{\circ}$, it is immediately obvious that a straight line path between those 3 weight data points would miss the true weights of the other inclinations. The other parameters sampled with only 3 grid points may also show this behaviour. For this particular scenario, the variation of the PCA weights with inclination may favour choosing a Matérn kernel for the emulator, as it allows less smooth functions than RBF. A further point of note is that other GP emulators designed on a similar architecture found Matérn kernels had a \textit{slightly} higher chance of outperforming other kernels \citep{kalmbach_estimating_2017, xu_standard_2025}. As a result, with \textsc{speculate}, the pre-trained emulators we distribute used the Matérn-5/2 kernel.  

A final point: when inspecting the GP diagnostics for the weight space `fit' in the \textsc{starfish} formalism, it is common to see the GP's higher-order principal components `miss' weight data points. It is important to note that this usually \textit{is not} the optimizer failing. Instead, each data point has the $\lambda_{\xi}$ noise term, which means the optimizer has the flexibility to fit the GP away from that data point. This term allows representation uncertainty in the projected weights. However, \textsc{speculate} includes a \textbf{strict weight fit} option which removes the $\lambda_{\xi}$ term. With this option, the GP fits directly in weight space and no longer `misses' weights. This option should generally be used only when diagnosing GP issues or when there is a high number of principal components, where the loss of variance-explained by the PCA is very small. 

\subsection{\textsc{speculate} inference}
\label{sec:inference}

Depending on the trained emulator method, two inference pathways are used in \textsc{speculate} to discover the physical parameters of a disc wind-dominated observation: a $\chi^2$ optimization for the regular-grid linear interpolation or neural network emulators, or a Bayesian inference engine following the \textsc{starfish} framework. Both pathways can impose prior constraints on the searchable parameter space, and as a result, both have access to a tool called the `parameter playground'. 

Emulators, by their nature, underperform as extrapolators. As the distance from the nearest grid points increases, the interpolating function becomes increasingly unconstrained and uncertain, as it lacks direction. Hence, by default, \textsc{speculate} imposes uniform priors within our \textsc{sirocco} grid bounds. For parameters expressed as logarithms, the prior is uniform in their logarithmic values. However, as mentioned in section~\ref{sec:Emu_GP}, the likelihood landscape is likely complex, so the prior limits set before inference are exposed to the user. Any additional physical information a user knows about a system can be incorporated to tighten the priors, aid the optimizer and break some possible parameter degeneracies. To give a concrete example: is the CV system being modelled eclipsing? If so, we know the inclination must be high ($>70^{\circ}$), and the mid to low inclination parameter space can be avoided. This can help break the degeneracy between inclination and disc accretion rate, as illustrated in the Tier 2 benchmark (section~\ref{sec:Tier2}).

This leads to the `parameter playground' tool. The playground lets a user manually search through the parameter space with the loaded emulator, essentially allowing pre-fitting to the data and the setting of additional prior bounds (if desired) before handed over to the inference optimizer\footnote{Any additional prior bounds used should be reported for reproducibility.}. These tools also let the user see in real time the individual effect a parameter has on a given set of spectral lines. This tool becomes very helpful once an inference result is achieved. Any parameter with a bimodal inferred posterior can be explored to compare the emulator's predicted spectra in the two regions of higher probability.

Beyond physical parameters, there is a selection of `nuisance' (non-\textsc{sirocco}) parameters (and priors) that can be configured. Largely taken from the \textsc{starfish} framework \citep{czekala_constructing_2015}, they consist of: 
\begin{itemize}
    \item \textbf{$A_V$ (Interstellar extinction in magnitudes)} -- This parameter can redden \textsc{sirocco} spectra to better match dust-affected observations. It uses the \textsc{extinction} package \citep{barbary_extinction_nodate} and the Fitzpatrick 1999 model \citep{fitzpatrick_correcting_1999} set to $R_V=3.1\,(=A_V/E(B-V))$, which corresponds to the UV/optical mean value for the diffuse ISM and defines the mean Galactic extinction law. 
    \item \textbf{Distance in pc ($d$) }-- This parameter, which is a wrapper around \textsc{starfish}'s natural-log scale ($\ln\,\mathrm{scale}$) parameter, directly scales the flux ($F$) according to the inverse square law so that $\ln\,\mathrm{scale} = 2\ln(100\,\mathrm{pc}/d)$, where $100\,\mathrm{pc}$ is our \textsc{sirocco} model observed distance. Distance is a highly degenerate parameter with other physical parameters that alter the continuum's flux level. As a result, a tight constraint is highly preferred for fitting, and \textsc{speculate} requires users to confirm their adopted distance and uncertainty via a checkbox before fitting with distance as a free parameter. To find CV distances, a good starting point is \textit{Gaia} parallax measurements. 
    \item \textbf{Continuum Tilt (1st order Chebyshev polynomial)} -- The overall shape of an observed spectrum can be difficult to calibrate accurately, so this parameter helps account for uncertainties in the continuum slope. This parameter, restricted so the 0th order (constant) term is 1, is a linear (gradient) continuum tilt applied to a spectrum. A positive value increases redder wavelengths, while a negative value suppresses them. 
    \item \textbf{GP Log Amplitude and Lengthscale ($\ln$; Gaussian Process emulators only)} -- These two parameters are global covariance residual terms of the Gaussian process kernel, which absorb some of the model-data mismatches. Randomly drawing from a GP covariance can yield many different realizations of the residual spectrum (model minus data). With an arbitrarily large number of samples, confidence bands of the ensemble's residuals can be calculated. If the fitted model minus data residual does not follow the ensemble's expected $1\sigma$, $2\sigma$, and $3\sigma$ limits, the log amplitude and lengthscale terms can be scaled appropriately to absorb this variance and prevent some parameter biases. These terms account for correlations between neighbouring residuals across the fitted spectrum, which can arise from observational effects and model mismatches. Further details can be found in section 2.3.1 of \citet{czekala_constructing_2015}.   
    \item \textbf{Local covariance matrix (Gaussian Process emulators only)} -- In addition to this global covariance, local kernels introduce the ability to add extra uncertainty around individual poorly modelled features whose residuals are not adequately described by the combined covariance ($C_{\mathrm{obs}}+C_{\mathrm{emu}}+C_{\mathrm{global}}$). This can help reduce their influence on the fit. We determine these local kernels from the residuals of an initial MLE fit and hold their positions, widths and amplitudes fixed during the subsequent MLE fit and MCMC sampling. Our current strategy for applying these local kernels is as follows: (i) We detect residuals beyond the $4\sigma$ limit of the combined covariance matrix along the diagonal. (ii) When an outlier is detected, the width of the region defining the local kernel (wavelength range) is grown until reaching a $2\sigma$ limit. This is to prevent other neighbouring pixels that might also exceed the $4\sigma$ limit from becoming separate local kernels. (iii) A Gaussian is fitted to the local region to determine a local kernel's mean (position), width (wavelength extent) and amplitude (how much additional variance is required to make the outlier a $3\,\sigma$ feature). (iv) A tapered local kernel is added using these values. Outside of the wavelength extent, no additional variance is added. These local kernels are often located at particular spectral lines where the model (\textsc{sirocco} in this case) cannot accurately represent the observation, for example, due to poor atomic data. However, with the current methodology, we aim to only down weight the influence of these lines on the inference fit, not just mask them out. This option may help prevent biased fitting.
\end{itemize}

\subsubsection{\texorpdfstring{$\chi^2$}{chi2} optimization}
\label{sec:chi-2 inference}
For the regular-grid linear interpolation and the neural network, \textsc{speculate} will only perform a maximum likelihood estimate using $\chi^2$ values. We have designed it this way for speed, simplicity, and familiarity. The first MLE optimization is initialized from the midpoints of the emulator grid bounds, except for the Nelder-Mead optimizer, whose simplex is set up to span (as best as possible) up to the edge of the parameter space. We have found that testing this Nelder-Mead simplex setup yields better results than the default behaviour, although we still recommend CMA-ES as the default optimizer. 

However, there is no guarantee that the optimizer has not fallen into a local minimum, particularly when there are many parameter degeneracies. As a result, there is an option to `Restart' the MLE a certain number of times and see whether the optimizer converges to a different solution. For the restart attempts, the starting positions are moved to random uniform sampled values inside the parameter bounds. For the $\chi^2$ optimization, restarts are fairly inexpensive, and several (3--5) are recommended. Once complete, a table of the overall best-fitting parameters is displayed, along with the best-fitting values for each restart. Dispersion between fits hints to the user that there are likely multiple solutions (a multimodal surface) for their observation, which complicates interpretation. A deeper look with a GP emulator/MCMC cornerplot is recommended to understand this solution landscape and uncertainties. However, if the MLE restarts are all fairly consistent, the user can act with more confidence and attempt modelling with \textsc{sirocco}. Each inference tool has an `Export .pf' button that automatically populates a \textsc{sirocco} input parameter file with the best-fitting values for the user to use with \textsc{sirocco}.

\subsubsection{\textsc{starfish} likelihood inference}
\label{sec:starfish inference}

A GP emulator follows a similar initial MLE stage, but instead uses \textsc{starfish}'s unnormalised flexible likelihood function (see equation 7 in \citet{czekala_constructing_2015}\footnote{the $\frac{N_{\mathrm{pix}}}{2}\,\ln(2\pi)$ term is constant and dropped}),

\begin{equation}
\label{equ:starfish likelihood}
    \ln\; \mathcal{L}= -\frac{1}{2}(R^TC_{\mathrm{total}}^{-1}R + \ln\;\det\;C_{\mathrm{total}})
\end{equation}

where $R$ is the model mean-data residual, and $C_\mathrm{total}$ is the composite covariance matrix.  This covariance matrix is constructed from a set of other additive covariance matrices, as described in equations 17 and 59 of \citet{czekala_constructing_2015}. This equation takes the form of

\begin{equation}
    C_\mathrm{total}=C_\mathrm{obs}+C_\mathrm{emu}+C_\mathrm{global} +\sum_k C_{\mathrm{local},k}.
\end{equation}
$C_\mathrm{obs}=\mathrm{diag}(\sigma_i^2)$ contains the observational flux variances. $C_\mathrm{emu}=X\Sigma_wX^\top$ encompasses the GP predictive covariance in PCA weight space transformed into flux space through 
\begin{equation}
    \Sigma_w=V_{22}-V_{21}\left[\Sigma_{\mathrm{grid}}+\frac{1}{\lambda_\xi}\left(\Phi^T\Phi\right)^{-1}\right]^{-1}V_{12},
\end{equation}
where crucially $\lambda_\xi$ acts as a precision parameter controlling the uncertainty on the PCA weights. Descriptions of the other terms can be found from equation 48 and 53 in \citet{czekala_constructing_2015}. $C_\mathrm{global}$ is the global covariance matrix describing correlated residuals across pixels, and $C_{\mathrm{local}}$ includes any additional local covariance matrices that reduce the influence of poorly represented individual features. 

In effect, this likelihood now has a better chance at accounting for model-observation mismatches resulting from an imperfect simulation, such as \textsc{sirocco}. In combination with this likelihood, the user can add additional prior information if they have further physical parameter constraints, turning the MLE into a maximum a posteriori setup. Also, several MLE restarts should be specified; the default is 3, but more restarts can help the optimizer search rugged likelihood/posterior spaces. Once the best-fitting MLE results are returned, the global best-fitting values are used as starting positions for the MCMC walkers, which map out the posterior distributions. 

Each parameter's walkers are initialized using a Gaussian ball centred on the MLE-estimated best-fitting value. For bounded uniform priors, the ball is truncated at the bounds, and the standard deviation is calculated as 15 per cent of the uniform prior width. For parameters that use a normal prior (typically \texttt{distance} and the GP log-amplitude), the same standard deviation as the normal prior is used for the Gaussian ball. Once evaluating, \textsc{speculate} is set up so that MCMC walkers that explore regions outside of our grid's parameter space are rejected outright. The \textsc{emcee} sampler \citep{foreman-mackey_emcee_2013} utilizes two walker movement strategies; (i) A `Differential Evolution' (DE) move \citep{nelson_run_2013} with 80 per cent probability, which helps handle linear correlations between parameters. (ii) A `Differential Evolution Snooker' (DESnooker) move \citep{ter_braak_differential_2008} with 20 per cent probability, which helps to escape local minima and traverse non-linear spaces. The \textsc{emcee} documentation highlights that this setup can perform better for lightly multimodal solutions. The user can then set the number of walkers, the number of steps and the number of burn-in steps ($b$). A minimum burn-in is calculated based on the per-parameter autocorrelation times \citep{foreman-mackey_emcee_2013} and if the user's minimum is less than that, the calculated step value is used instead, following $b=\max(b_{\mathrm{user}}, 2\max\tau_k)$. For \textsc{speculate}, the exploratory defaults are set at $64$ walkers, with $2\,500\text{--}3\,000$ steps, and a $500$ step burn-in. If a walker steps outside the emulator grid, the proposed position is rejected before evaluation. Ultimately, the same broad MLE-to-MCMC workflow is applied to our Tier 2 parameter retrieval benchmark and our Tier 3 application-to-observations benchmark, which are discussed later in sections~\ref{sec:Tier2} and \ref{sec:Tier3}.  

\subsection{Accelerating the \textsc{starfish} framework}
\label{sec:accelerate_starfish}

Nevertheless, even with the previously detailed changes, training a GP on tens of PCA components with thousands of grid points remains computationally demanding. The bulk of \textsc{speculate}'s engineering has gone into refactoring parts of \textsc{starfish}'s code to accelerate the computational bottlenecks. We have already detailed the block-diagonalization to reduce the memory complexity and overall computational time by requiring fewer calculations in section~\ref{sec:The computational scaling of the PCA--GP}. However, we have an additional 4 refactors that significantly accelerate the \textsc{starfish} framework for higher dimensions. 

Firstly, and likely our biggest change, is the transition from a pure CPU implementation to a CPU-GPU approach with PyTorch and RAPIDS's cuML. The key use here is computing the expensive Cholesky decomposition on the GPU, along with other matrix operations involved in a call to the likelihood function, and the PCA decomposition. While the CPU path remains a fallback approach, using a CPU beyond $500\text{--}1000$ grid points is a futile endeavour. Now, with our GPU implementation, we \textit{could} technically utilize all $78\,732$ grid points of the \textsc{speculate}'s boundary layer emission CV grid (provided a user has access to a top-of-the-range GPU with up to $96\,GB$ VRAM like an Nvidia RTX PRO 6000 Blackwell Series) \footnote{It should be noted that it is extremely unlikely that a user will use \textsc{speculate} in this fashion. Degeneracy between this many uninformed parameters will lead to poor results, and a user should inform the prior space for \textit{some} parameters from knowledge of the system or previous smaller grid space inferences}. However, because of the caching inference steps we will detail in the next paragraph, there is a more practical computational time-scale limit of $30\,000$ grid points for these cards, and a rough limit of $5000$ grid points for entry-level GPU cards with $6\,GB$ of VRAM. It should be noted here that our no boundary layer emission CV grid with a mid-level sampling of inclinations can fit comfortably on these entry-level cards \footnote{Analysing the April 2026 Steam Hardware \& software survey showed $\approx81$ per cent of GPUs have $6\,GB$ of VRAM or higher} with $4374$ grid points. 

Secondly, as hinted at previously, we are cautious about the memory complexity trade-off to the computational time complexity. As a result, we utilize the maximum available memory to store repeated values in calculations before running out of memory and having to recalculate the same maths repeatedly with every call to the emulator. There are two important points where this is the case: (i) In the training, if the full block diagonal $V_{11}$ matrix is too large for the GPU memory, the training routines will calculate each block on the fly instead of storing the full matrix. This means emulators with a higher number of grids likely train slower than if all their blocks could (theoretically) fit into GPU memory. However, in practical terms, this means we can train on tens of thousands more grid points while keeping computational time reasonable (on the order of hours). (ii) In the inference, we cache per-component Cholesky factors and pre-computed GP solve terms in VRAM. This means we can reduce the MLE and MCMC evaluations to a single expensive call and repeated (and cheaper) triangular solves. If the memory cannot hold the required cached items, then the terms are recomputed at each evaluation. However, in practice, the size of grids that would be out of memory would take too long to recompute thousands of times. This is the point that meant, in the previous paragraph, we were currently limited in practical terms to $20\,000\text{--}30\,000$ grid points on the best hardware, as the cache size of these items exceeds $96\,GB$ VRAM.

Thirdly, the \textsc{starfish} framework was designed to be flexible, enabling the processing of irregular grids for emulator construction. However, this meant they required \textsc{scipy}'s `LinearNDInterpolator' function \citep{virtanen_scipy_2020} to interpolate between scattered data points for the stored per-spectrum normalization factors, which is incredibly slow. Instead, we designed our \textsc{speculate} grids to be quasi-regular (a rectilinear Cartesian product), allowing us to use \textsc{scipy}'s `RegularGridInterpolator' function \citep{virtanen_scipy_2020}, which is near instantaneous (millions of times faster) for high numbers of grid points. As a result, \textsc{speculate} requires that any new \textsc{sirocco} grids added to the program be regularly gridded. 

Finally, we introduce early stopping/convergence tolerance procedures for both the emulator training and MLE inference stages to prevent unnecessary computation. After a short burn-in period that initializes the negative log-likelihood to sensible values for a given configuration, a tolerance level is calibrated, depending on the method and the likelihood values encountered, to prevent long, drawn-out, and unnecessary loss curves. Additionally, when repeated MLE inference evaluations of the negative log-likelihood values are essentially flat, the inference stops. Hard caps can also be set for a maximum number of iterations/evaluations. Additionally, during training, if a user is unsure whether an emulator has fully converged, the user can \textbf{continue training} the emulator, and \textsc{speculate} will begin from the corresponding pre-trained emulator. This also allows the user to refine the emulator if desired, as detailed in section~\ref{sec:Emu_GP}.

\section{Emulator performance benchmarks}
\label{sec:Emulator Performance Benchmarks}

\begin{table*}
\centering
\footnotesize 
    \setlength{\tabcolsep}{3.5pt} 
    \begin{tabular}{l c c c}
        \hline
        \textbf{Emulator Configuration} & \textbf{UV} & \textbf{Blue-Optical} & \textbf{Red-Optical} \\
        \hline
        \textbf{Architectures} & GI, NN, GP & GI, NN, GP & GI, NN, GP \\
        \textbf{Grid} & \multicolumn{3}{c}{\texttt{speculate\_cv\_no-bl\_grid\_v87f}} \\
        \textbf{Inclination Sampling, GI/NN (°)}& Full (30--85; $\Delta i=5$) & Full (30--85; $\Delta i=5$) & Full (30--85; $\Delta i=5$) \\
        \textbf{Inclination Sampling, GP (°)}& Mid (30--80; $\Delta i=10$) & Mid (30--80; $\Delta i=10$) & Mid (30--80; $\Delta i=10$) \\
        \textbf{Wavelength Range (Å)} & 1150--1750 & 3400--5250 & 5750--6850 \\
        \textbf{PCA Components} & 30 & 30 & 30 \\
        \textbf{PCA Variance-Explained} & 96.672\% & 98.542\% & 97.693\% \\
        \textbf{Flux Scale} & Linear & Linear & Linear \\
        \textbf{Smooth Input Data?} & Yes & Yes & Yes \\
        \textbf{GP Kernel} & Matérn-5/2 & Matérn-5/2 & Matérn-5/2 \\
        \hline 
    \end{tabular}
\caption{The example emulator configurations used for inference and performance benchmarks throughout this paper. All emulators trained above use the \texttt{linear} flux scale which is \textsc{sirocco}'s spectral flux in erg\,s$^{-1}$\,cm$^{-2}$\,Å$^{-1}$ at $100\,\mathrm{pc}$.}
\label{tab:emulator_config}
\end{table*}

At this point, the sampling speed of a trained emulator is fast. Individual inference evaluations of the GI and NN methods typically will take no longer than a few milliseconds for most CPUs. The GP takes on the order of tens of milliseconds on GPU\footnote{On an Intel Xeon Gold 6338 CPU and Nvidia A100 GPU with CUDA 12.8, a GI sample takes $2.43\,\mathrm{ms}$, a NN sample takes $1.66\,\mathrm{ms}$, and a GP sample takes $37.48\,\mathrm{ms}$}. However, as \textsc{speculate} is an emulator-based inference framework, there are a large number of possible emulator configurations. As a result, this paper cannot simply quote some statistics describing its overall performance. Performance will change depending on a researcher's individual configuration. For example, what is the emulator wavelength range? How many spectral lines are being fitted simultaneously? How many PCA components are included? Is the \textsc{sirocco} data smoothed? What prior information is known about the user's system? What physical parameters are included in the emulator? What flux space are we emulating in? And so on. Hence, quoting a set of statistics will be irrelevant for most users. 

Instead, we expose benchmark tools so that users can evaluate the performance of their own emulators. \textsc{speculate}'s benchmark process has three tiers, each of which builds on the previous, and they are of increasing difficulty. These are: 
\begin{itemize}
    \item \textbf{Tier 1 -- Grid Reconstruction}: How well does an emulator reproduce its own training and test grid spectra when sampling their exact parameter combination?
    \item \textbf{Tier 2 -- Parameter Recovery}: How well can our emulator inference recover the parameters of \textsc{sirocco} test grid spectra when searching within an input grid's parameter space bounds?
    \item \textbf{Tier 3 -- Modelling Observational Spectra}: How well can our emulator-based inference framework fit an actual \textsc{sirocco} model to an observational spectrum?    
\end{itemize}

To guide readers through the benchmark process, we will use three pre-release emulators, whose full configurations are provided in Table~\ref{tab:emulator_config}. The \textbf{UV} emulator covering $1150\text{--}1750\,$Å includes spectral lines such as \ion{C}{iii}, Ly\,$\alpha$, \ion{N}{v}, \ion{O}{v}, \ion{Si}{iv}, \ion{C}{iv}, and \ion{He}{ii}.  The \textbf{Blue-optical} emulator covering $3400\text{--}5250\,$Å includes spectral lines/regions such as the Balmer Jump, H\,$\epsilon$, H\,$\delta$, H\,$\gamma$, the Bowen blend of \ion{C}{iii} and \ion{N}{iii}, \ion{He}{ii}, H\,$\beta$ and \ion{He}{i}. The \textbf{Red-optical} emulator covering $5750\text{--}6850\,$Å is more limited on spectral lines but includes \ion{He}{i} emission lines and H\,$\alpha$\footnote{Naturally, some ion names appear in more than one region as they correspond to different atomic transitions.}. These emulators are designed to be robust but push the limits to reveal where users may have difficulty fitting \textsc{sirocco} models to their observations. 

For the Tier 1 and 2 benchmark, we need \textit{unseen} \textsc{sirocco} models. For this, each CV grid released also includes a corresponding test grid of 50 models, with 12 inclination angles, for 600 spectra in total. For the Tier 3 comparison, we need observations. Hence, \textsc{speculate} ships with 5 different observational CV spectra for users to test with: IX Vel, RW Sex, RW Tri (de-reddened), UX UMa, and V3885 Sgr. The distances and literature inclinations of these 5 systems are listed in Table~\ref{tab:distances_and_inclinations}, and they span a wavelength range of $\approx1\,150\text{--}10\,250\,\mathring{A}$, covering nearly the entire UV and optical spectrum. We (the \textsc{sirocco} collaboration) have previously studied these systems using \textsc{sirocco} (formerly known as \textsc{python} before its renaming), and readers should consult \citet{matthews_impact_2015} for the more traditional fitting approach we used before \textsc{speculate}. 

\subsection{Tier 1 -- Grid reconstruction performance}
\label{sec:Tier1}

Having defined the emulator architecture, we now quantify how well a trained \textsc{speculate} emulator reproduces the training and test grids. No machine learning model is ever perfect. However, if performance here is poor, the emulator should not be used in any subsequent inference steps. There are a few general places a \textsc{speculate} emulator can fall over here. Firstly, modelling too many spectral lines with too few principal components. A setup with too small a variance-explained score will naturally lead to a high baseline PCA approximation error in absolute terms. Secondly, incomplete/poor training of an emulator, typically achieved by hitting the maximum number of iterations set by the user. If a loss curve has not converged (asymptoted), this typically results in the selected architecture producing excessive residual error. 

\begin{figure*}
\centering

	\includegraphics[width=\linewidth,height=0.74\textheight,keepaspectratio,
        alt={Three panels plot reconstruction RMSE against wavelength for the ultraviolet, blue-optical and red-optical test spectra. Errors increase around spectral features. The three emulator methods show broadly similar performance above the PCA truncation curve.}
    ]{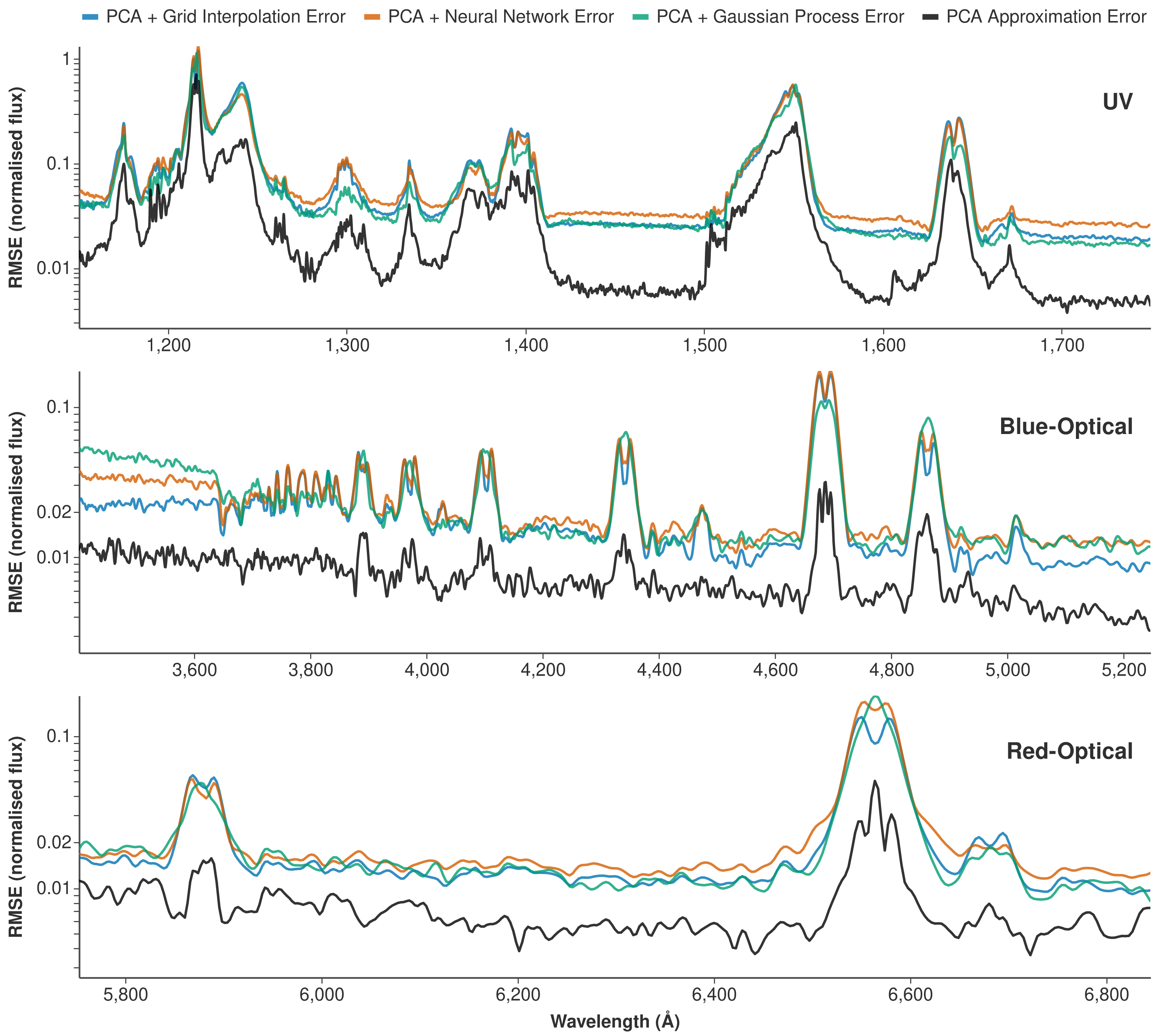}
    \caption{The per-wavelength test-grid flux RMSE for the UV, blue-optical, and red-optical emulator configurations given in Table~\ref{tab:emulator_config}. Each curve is calculated from $550$ reconstructed spectrum residuals from the 50 test-grid \textsc{sirocco} models across 11 inclinations ($30\text{--}80^\circ$). The black line shows the reconstruction error obtained from a PCA truncation to 30 principal components. The blue, orange, and green line shows the total reconstruction error for the grid interpolation method, the neural network, and the Gaussian Process, respectively. All spectra are mean normalized before RMSE calculations.}
    \label{fig:Test-RMSE-Tier1}
\end{figure*}

The key way a user can see whether an emulator performs poorly on its own data is by plotting the per-wavelength RMSE (root mean square error) envelope. We have two flavours of this plot in Tier 1: one evaluating against training-data grid points, and another evaluating against the test grid. While the training grid version is presented in Appendix~\ref{sec:Tier 1-- Training grid RMSE envelopes}\footnote{Although all outputs are in RMSE and can be showcased on one plot, the methodologies differ enough that between architectures' performance should not be compared based upon this plot.}, in this section, we present the test grid version. Fig.~\ref{fig:Test-RMSE-Tier1} shows the per-wavelength test grid RMSE envelopes across our emulator configurations. Here, the curves are calculated from the reconstructed (emulated) spectrum residuals at the grid points of our 50 \textsc{sirocco} test-grid models across 11 inclinations from $30^{\circ}$ to $80^{\circ}$ in $5^{\circ}$ increments. Before each RMSE is calculated across the 550 spectra, each true and emulated spectrum is mean normalized using the same pre-processing as the \textsc{speculate} emulators. The RMSE is calculated through the following equation: 
\begin{equation}
    \mathrm{RMSE}(\lambda) = \sqrt{\frac{1}{N}\sum_{i=1}^N\left[f_i^{\mathrm{sirocco}}(\lambda)-\hat{f}_i^{\mathrm{emulator}}(\lambda) \right]^2}
\end{equation}
where $N$ is the number of test grid spectra evaluated, and $f$ is the flux in the mean-normalized space, where each input spectrum is divided by its mean flux across the selected wavelength range. As each spectrum is normalized to have a mean flux of one, an RMSE approaching unity corresponds to a RMS residual of the same scale as the mean flux. 

From Fig.~\ref{fig:Test-RMSE-Tier1}, we can see that all architectures show broadly the same reconstruction performance among our emulator configurations. No method performs universally better across the wavelength ranges. The majority of errors contributed to the forms around the spectral features (as would be expected), emphasizing that achieving a more balanced PCA error contribution is more important than an individual architecture method alone. Architectural choices should instead be based on researchers' needs. For example, a grid interpolation approach is essential if training and inference speed are critical. The neural network approach, with non-linear mappings, offers an advantage for PCA weight interpolation, or for the Gaussian process if the inclusion of an emulator's (estimated) uncertainty in the posteriors is required. Beyond these RMSE plots, Tier 1 includes a suite of architecture-specific additional diagnostics, which we will discuss below.  

Let us start with the additional diagnostics for GI and NN emulators, where all their diagnostics are self-contained in the `Quick Fit Notebook'. \textsc{speculate} reports $R^2$ scores for how well these emulators can explain the variance in their input data PCA weight space, in the flux space input spectra, and in the flux space of the corresponding unseen test grid spectra (spectra away from the regularly gridded points). For the GI method, the regular-grid PCA-weighted $R^2$ and regular-grid flux $R^2$ scores use an approximate leave-one-out (LOO) method to score each data point. The GI attempts to predict it using the mean of its nearest neighbours; otherwise, the reported $R^2$ would always be perfect. For the NN, the network does not necessarily have to pass through weight data points. Therefore, we do not need to remove any data points in its weight predictions.

Three $R^2$ values are given for each method: the mean $R^2$ score, the 5th percentile $R^2$ score, and the worst $R^2$ score. An example of what these scores may look like for our trained UV GI emulator can be seen in Table~\ref{tab:UV_GI_R2_diagnostics}. A strong emulator will have a reasonably consistent $R^2$ score across spectra. We choose the 5th percentile and the worst $R^2$ score to see if there is a sudden drop-off in performance. We can see an example of this in our table where the regular-grid flux fifth-percentile-to-worst scores drop by 0.416 compared to the corresponding test grid's difference of 0.179. Such a drop can flag a small number of spectra that are particularly hard to emulate given the input data. If the mean score is low, then a user may need to focus on emulator performance with a higher number of components and/or restrict the emulator wavelength range. A user should also be aware of the possibility of creating `fake' high $R^2$ scores when an emulator is configured over a very wide wavelength range with minimal features. If the continuum shape dominates the variance across wavelength, an emulator could match it well while poorly reproducing spectral features and still report a high flux $R^2$. Similarly, a high PCA variance-explained score does not guarantee accurate reconstruction of every spectral feature. 

\begin{table}
\centering
\small
\setlength{\tabcolsep}{3pt}
\begin{tabular}{lccc}
        \hline
        \textbf{Diagnostic} & \textbf{Mean} & \textbf{5th percentile} & \textbf{Minimum} \\
        \hline
        Regular-grid PCA-weight $R^2$ & 0.725 & 0.560 & 0.516 \\
        Regular-grid flux $R^2$       & 0.949 & 0.829 & 0.413 \\
        Test-grid flux $R^2$          & 0.874 & 0.776 & 0.597 \\
        \hline
    \end{tabular}
\caption{Example $R^2$ score diagnostics for the UV grid-interpolator emulator configuration given in Table~\ref{tab:emulator_config}. The regular-grid diagnostics use approximate leave-one-out predictions. The PCA-weight statistics are calculated across the 30 components, and the flux statistics are calculated across 500 randomly selected regular-grid spectra and 600 test-grid spectra, respectively.}
\label{tab:UV_GI_R2_diagnostics}
\end{table}

We also include a smorgasbord of additional diagnostics for GP emulators. While the number of diagnostics is too extensive to discuss in detail in this section, descriptions of what each plot shows, what `better' looks like, what to watch for and how to improve performance are provided with all plots in the Tier 1 benchmark. Firstly, our LOO RMSE training grid score, shown in Fig.~\ref{fig:train-RMSE-Tier1}, is broken down by component contribution, visualized as a bar chart to identify which components are sources of large error, along with per-component $R^2$ scores. If a particular component is causing issues with the emulator's performance, this should be highlighted in these diagnostics and indicate how well the GP's predictive mean is performing. Additionally, the negative log predictive density per component is plotted to assess whether the GP's uncertainties (variance) are well calibrated. There are LOO residual distribution histograms, Q-Q plots, and decorrelated Q-Q plots of the LOO residuals to indicate if and where an emulator is under- or overconfident. Tier 1 also hosts an interactive reconstructed spectra explorer to view the performance of individual reconstructions and the strength of uncertainty across different wavelengths, along with plots of each eigenspectrum.

\subsection{Tier 2 -- Parameter retrieval performance}
\label{sec:Tier2}

Testing the reconstruction accuracy of the grids is necessary to determine whether a trained emulator can reproduce our \textsc{sirocco} spectra. However, it is not a sufficient test to determine whether our emulator inference engine is truly useful and able to recover the physical parameters of an \textit{unseen} \textsc{sirocco} spectrum. This is what the Tier 2 benchmark tests. Here, 600 CV test grid spectra (50 models at 12 different inclinations) are available to pass through \textsc{speculate}'s inference workflow. We can then measure the accuracy of the emulator's best-fits to each spectrum and compare the emulator's reported best-fitting parameters with the ground-truth \textsc{sirocco} parameters. All these test grid models are Latin hypercube samples from within the bounds of the matching training grid's parameter space.

For this test, ideally, a user would pass a trained emulator through the Tier 2 inference stages for all 50 models. However, for \textsc{speculate}'s largest emulators' configurations (all parameters included), this could take up to a day per model with entry-level GPUs such as Nvidia's 1000 or 2000 series cards. More modern-day consumer GPUs, such as 5000- and 6000-series cards, or server-grade GPUs like A100, H100, or B100 cards, will be an order of magnitude faster. However, in practical terms, depending on hardware availability, a smaller number of models might be required, for example, $5\text{--}10$ models. 

During an inference, there are options to freeze parameters (both physical and nuisance) to prevent the optimizer from searching that corresponding parameter space. In Tier 2, several of these are ticked by default. Firstly, we know that the \textsc{sirocco} spectra are observed at 100\,pc with no extinction. As our emulator is also trained on \textsc{sirocco} data, we confidently freeze the distance and $A_V$ parameters. Additionally, we do not have any systematics that might affect the continuum slope of the spectrum so that we can freeze the \texttt{cheb\_1} term as well (the Chebyshev polynomial correction in \textsc{starfish}). The remaining nuisance parameters, the GP global covariance amplitude and length scale, are left thawed and marginalized over in the MCMC stage to determine the final physical parameter values. 

\begin{figure*}
\centering
    \includegraphics[width=\linewidth,height=0.74\textheight,keepaspectratio,
        alt={The upper panel compares the input ultraviolet spectrum with emulator predicted spectra at the true parameter value, from the maximum-likelihood fit and the spectrum at the MCMC median parameters. The lower panel shows residuals relative to the input, with the largest discrepancies around strong spectral lines.}
    ]{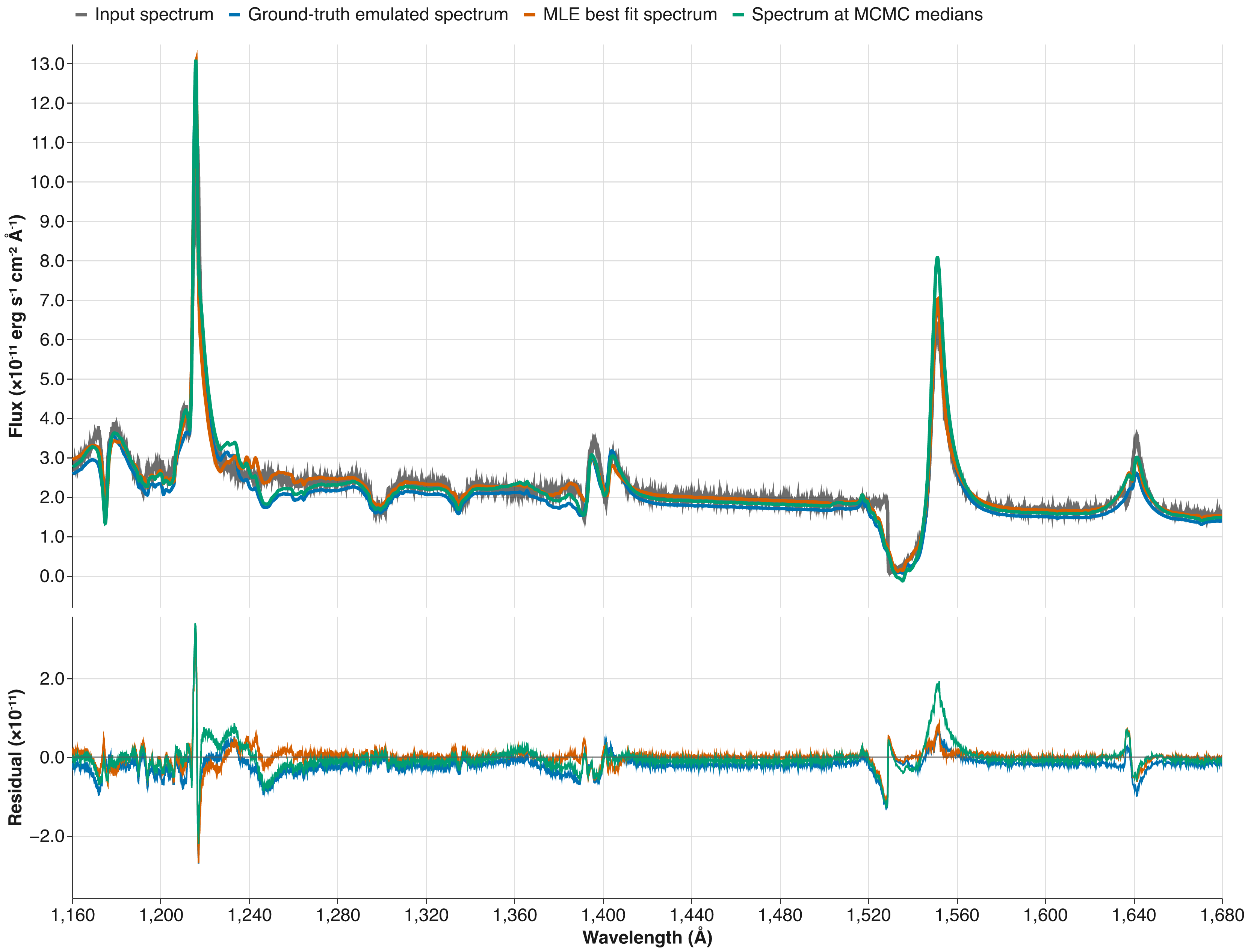}
    \caption{Best-fits of \textsc{speculate}'s MLE (in orange) and MCMC (in green) inference steps for \textsc{sirocco} test grid model run34 at $45^{\circ}$ for the \textbf{UV} emulator configuration given in Table~\ref{tab:emulator_config}. The corresponding best-fitting parameter values are shown in Table~\ref{tab:run34_results}. A comparison of the emulated spectrum using the ground-truth physical parameter values is shown in blue. The \textsc{sirocco} input spectrum is shown in grey. The top panel shows the spectra, while the bottom panel shows the residuals of the spectra away from the input spectrum.}
    \label{fig:run34_fits}
\end{figure*}

\begin{table*}
\centering
\small
\setlength{\tabcolsep}{3pt}
\begin{tabular}{l l l c c c c c c}
        \hline
        \makecell{\textbf{Physical}\\\textbf{Parameter}} & \textbf{Symbol} & \textbf{Unit} & \makecell{\textbf{Ground}\\\textbf{Truth}} & \makecell{\textbf{MLE}\\\textbf{Fit}} & \makecell{\textbf{MLE}\\\textbf{$\Delta$}} & \makecell{\textbf{MCMC}\\\textbf{Fit}} & \makecell{\textbf{MCMC}\\\textbf{$\sigma$}} & \makecell{\textbf{MCMC}\\\textbf{$\Delta\sigma$}} \\
        \hline
         Disc Accretion Rate & $\log_{10}(\dot{M}_{\mathrm{disc}})$ & dex & $-7.8447$& $-7.8152$& +0.0295& $-7.8029$& 0.0736& $+0.57\,\sigma$\\
         Wind Mass Loss Rate & $\dot{M}_{\mathrm{wind}}$ & $\dot{M}_{\mathrm{disc}}$& 0.2634 & 0.1366& $-0.1268$& 0.2311& 0.0549& $-0.59\,\sigma$\\
         Wind Collimation & $\log_{10}(d)$ & dex& 0.8479 & 0.6861& $-0.1618$& 0.9696& 0.2844& $+0.43\,\sigma$\\
         Mass Loss Rate Exponent & $\alpha$ & -- & 0.9156 & 0.9949& +0.0793& 0.8629& 0.1142& $-0.46\,\sigma$\\
         Acceleration Length Scale & $\log_{10}(R_{\mathrm{s}})$ & dex & 10.3332 & 10.3497& +0.0165& 10.5134& 0.2055& $+0.88\,\sigma$\\
         Acceleration Exponent & $\beta$ & --& 3.7790 & 4.4618& 0.6828& 3.1186& 0.9373&$-0.70\,\sigma$\\
         Inclination & $i$ & $^{\circ}$& 45.0000 & 44.9967& $-0.0033$& 45.7824& 5.3123& $+0.15\,\sigma$\\
        \hline
    \end{tabular}
\caption{Best-fitting parameter values through \textsc{speculate}'s inference steps for \textsc{sirocco} test grid model run34 at $45^{\circ}$ for the \textbf{UV} emulator configuration given in Table~\ref{tab:emulator_config}. The corresponding best-fitting plots are shown in Fig.~\ref{fig:run34_fits}. The MCMC corner plot is shown in Fig.~\ref{fig:run34_corner} and uses median values as best-fitting values.}
\label{tab:run34_results}
\end{table*}

\begin{figure*}
\centering
    \includegraphics[width=\linewidth,height=0.85\textheight,keepaspectratio,
        alt={A corner plot that shows the one- and two-dimensional marginal posterior distributions of seven wind parameters for test model run34. Dashed lines mark posterior summaries and red lines show the true parameter values; contours show correlations and broad or multiple peaks.}
    ]{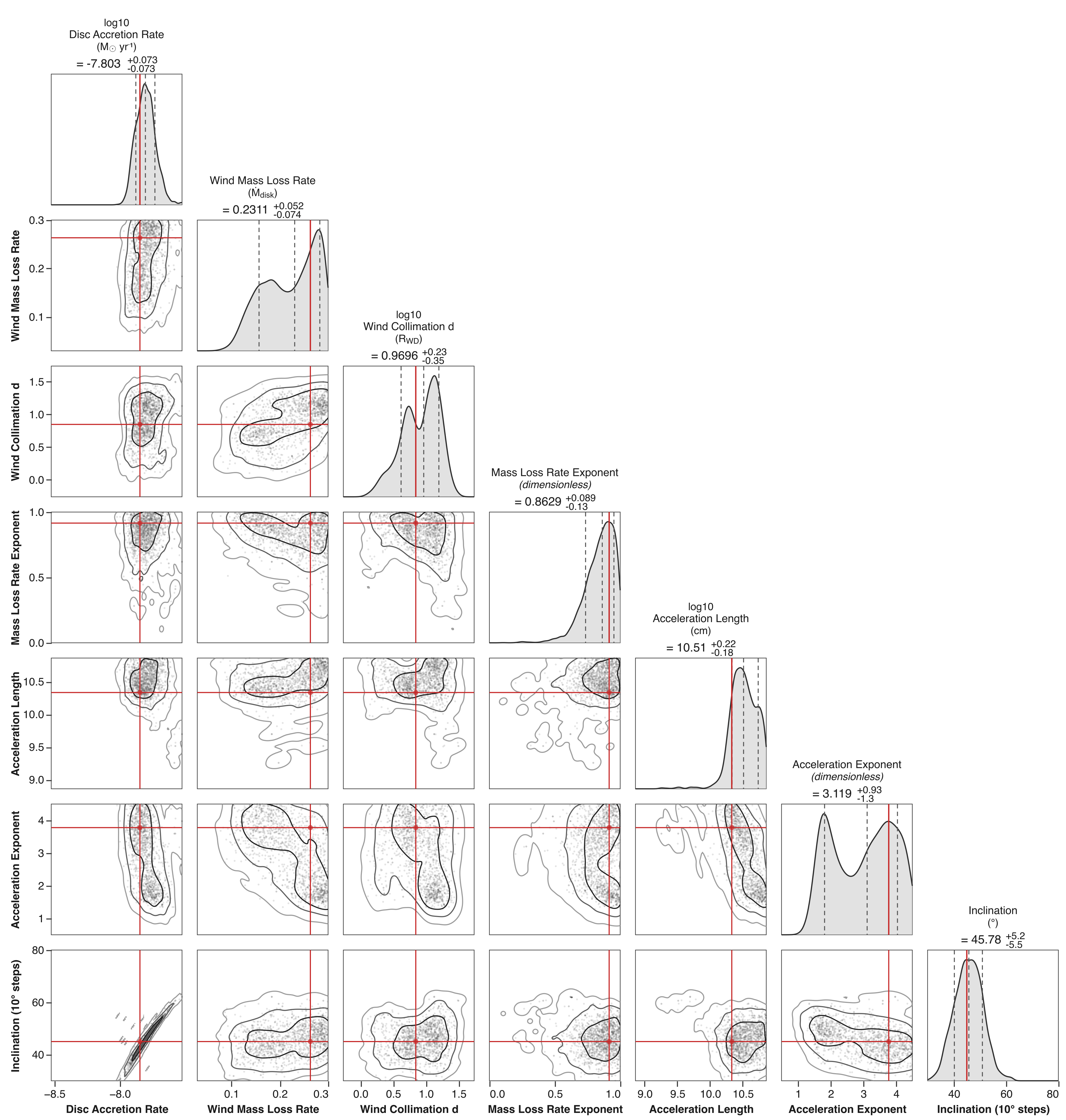}
    \caption{An MCMC corner plot showing the marginalized 1D and 2D posterior probability distributions for run34's \textsc{sirocco} model parameters. The 2-D plot contours enclose 68.27 per cent, 95.45 per cent, and 99.73 per cent of each pairwise marginal posterior. The vertical black dashed lines highlight the median and the $1\,\sigma$ credible interval of each diagonal 1-D probability density function. The red lines show the ground-truth parameter values of run34. Each corner plot's axis scale equals the bounded prior space. Strong correlations between parameters can indicate degeneracies. Separate peaks may indicate multimodality, with multiple possible solutions to a single fit.}
    \label{fig:run34_corner}
\end{figure*}

To demonstrate what is happening behind the scenes for a Tier 2 benchmark, we are going to show explicitly one model inference first. For this, we will use the test grid model run 34 at $45^{\circ}$ in the UV. We have chosen this model for two reasons. Firstly, the model exhibits a variety of spectral line strengths and shapes. There are small absorption features, strong emission features, an obvious \ion{C}{iv} P-Cygni profile and an apparent double-peaked line. Secondly, the ground-truth physical parameters for the model are moderately distant from all grid points, meaning we are not fitting a model that is conveniently near a seen model during training. We fit the model using our UV emulator configuration, with 3 MLE restarts and 3000 MCMC steps, including a 500-step burn-in. The best-fitting values can be seen in Table~\ref{tab:run34_results}, their corresponding spectra in Fig.~\ref{fig:run34_fits} and the MCMC cornerplot in Fig.~\ref{fig:run34_corner}. \textsc{speculate}'s inference for run34 is strong with all best-fitting parameter values within $1\sigma$ of the true parameter values. Notably, for several parameters, the MCMC yielded median results closer to the ground truth than the MLE results. This result provides evidence of the value of the MCMC stages and demonstrates that MLE values should serve only as a starting point for the overall inference. 

We can now scale this methodology to all 50 test grid models with round-robin randomised inclinations and view the reported coverage probability\footnote{Note: Our UV-trained emulator only covers inclinations between $30^{\circ}$ and $80^{\circ}$, hence, we remove $85^{\circ}$ inclined spectra from the rotation otherwise our emulator will be attempting to predict out of grid bounds (extrapolating).}. The coverage probability is the percentage of times the true parameter value falls within a particular confidence interval ($1\sigma$, $2\sigma$, or $3\sigma$) and measures how reliable our estimates are. An optimal result would be for all parameters' coverage probabilities to be approximately the reported $\sigma$ level. Table~\ref{tab:Tier2_coverage_probability} reports that our coverage probabilities for this particular UV emulator are \textit{less} than our expected $\sigma$ levels. This means our reported errors are \textit{overconfident}, with particular weaknesses in the inclination parameter, which also has a knock-on effect on the coverage of other parameters through its degeneracies. Incomplete MCMC sampling of the posterior may also contribute to the undercoverage reported here, where individual fits may require longer chain lengths. 

From testing performance over several emulator configurations, for emulators up to \textsc{speculate} v0.5.0, this overconfidence tends to be a feature rather than a one-off. Improving the calibration accuracy of these errors is still a work in progress. However, users should note this particularly strong sensitivity to inclination. This is the weakest point, as there are strong degeneracies between other physical parameters and inclination. A user should try hard to train with mid to full level inclination sampling and/or restrict the range of possible inclinations using independent constraints where available. An estimate of inclination, even if only low, median, or high, can help improve these coverages. Along with this point, distance is another strongly degenerate parameter among the physical parameters, as distance, disc accretion rate, wind mass-loss rate, collimation, and inclination all affect the continuum flux level. For CVs, users should have a fairly concrete distance estimate thanks to \textit{Gaia}, and prior ranges should be set tightly to help break some of these degeneracies. If the flux scale is unknown or poorly constrained, configuring an emulator to use a \texttt{continuum-normalized} flux scale should be the default.

Whilst not necessary to present here, each Tier 2 benchmark completed is accompanied by shrinkage scores, probability calibration curves (PP-plot), and per-parameter RMSE and bias scores for diagnostic purposes. The shrinkage scores indicate the percentage by which the posterior space shrinks from the original prior space for each parameter. Naturally, the larger the reduction, the better, as long as the coverage probability remains sensible. The PP-plot will indicate whether this is true. Ideally, each parameter will sit along the $y=x$ line, where, say, a 68 per cent credible interval will encapsulate the parameter 68 per cent of the time. However, as noted earlier, our emulators tend to be overconfident up to version 0.5.0, so parameter lines are generally below the $y=x$ line. This leads to the per RMSE and bias scores, which should highlight parameters of particular concern. Like Tier 1 plots, the Tier 2 plots have accompanying notes on what to be aware of for each plot within \textsc{speculate}.

\begin{table}
\centering
\small
\setlength{\tabcolsep}{3pt}
\begin{tabular}{l l l l}
\hline
        \multicolumn{1}{c}{\textbf{UV Emulator Configuration}} & \multicolumn{3}{c}{\textbf{Coverage Probability}} \\
        \hline
        \textbf{Physical Parameter} & \textbf{$1\,\sigma$} &  \textbf{$2\,\sigma$} & \textbf{$3\,\sigma$} \\
        \hline
        Disc Accretion Rate  &42\% &	84\% &	98\% \\
        Wind Mass Loss Rate &52\% &	88\% &	100\% \\
        Wind Collimation &40\% &	78\% &	88\% \\
        Mass Loss Rate Exponent &46\% &	88\% &	94\% \\
        Acceleration Length Scale  &46\% &	82\% &	94\% \\
        Acceleration Exponent  &48\% &	84\% &	92\% \\
        Inclination  &26\% &64\% &  70\% \\
        \hline
    \end{tabular}
\caption{Coverage probabilities of the UV emulator for each physical parameter over 50 non-boundary layer CV test grid spectra, resulting in values reported to 2 per cent intervals. Coverage is evaluated using central equal-tailed posterior intervals. }
\label{tab:Tier2_coverage_probability}
\end{table}

\subsection{Tier 3 -- Application to CV observations}
\label{sec:Tier3}

The Tier 1 and Tier 2 benchmarks assess \textsc{speculate}'s ability to reproduce \textsc{sirocco} spectra and recover their input parameters. At this point, the next question becomes more interesting: how well can a \textsc{sirocco} spectrum, within \textsc{speculate}'s synthetic library grid bounds, represent a user's actual observation? This is our Tier 3 benchmark and the most applicable benchmark for observations. Tier 3 is essentially the workflow we would expect a user to follow when using \textsc{speculate}'s inference tool, but for 5 preloaded demonstration \textit{HST} observational spectra included with \textsc{speculate}. These spectra were provided by Knox Long and obtained with \textit{HST/STIS} as part of programme \textit{14637} (PI: Long), which observed five nova-like variables across approximately $1150\text{--}10\,250\,$Å. The observations used the E140M, G140L, G230LB, G430L and G750L gratings. For each target, while the different wavelength bands were not observed simultaneously, they were observed within a single \textit{HST} orbit, helping to limit the effects of variability when combining the spectra \citep{hubeny_improved_2021}. This workflow is as follows: the nuisance parameters ($A_V$, distance, \texttt{cheb\_1}, GP covariance amplitude and lengthscale) are treated as free parameters in MLE and then marginalized over in MCMC. Distance values and inclination priors are added\footnote{For the 5 observations included, these values are preloaded in the Tier 3 benchmarks code.}. The resultant MCMC physical parameters are then exported to a sirocco parameter input file. A full \textsc{sirocco} simulation is run using this parameter file, and the resultant spectra are plotted with the corresponding posterior median nuisance parameter transformations included against the observation. This comparison is the ultimate test of whether \textsc{speculate} can be used reliably to find a \textsc{sirocco} spectrum that is a decent match to data. 

\begin{figure*}
\centering
    \includegraphics[width=\linewidth,height=0.74\textheight,keepaspectratio,
        alt={Three panels compare observed spectra with the emulator predicted spectrum and matching SIROCCO simulated spectrum at the posterior median parameters: V3885 Sgr in the ultraviolet, RW Tri in the blue optical and UX UMa in the red optical. Several lines agree, but discrepancies remain in some absorption edges and emission peaks.}
    ]{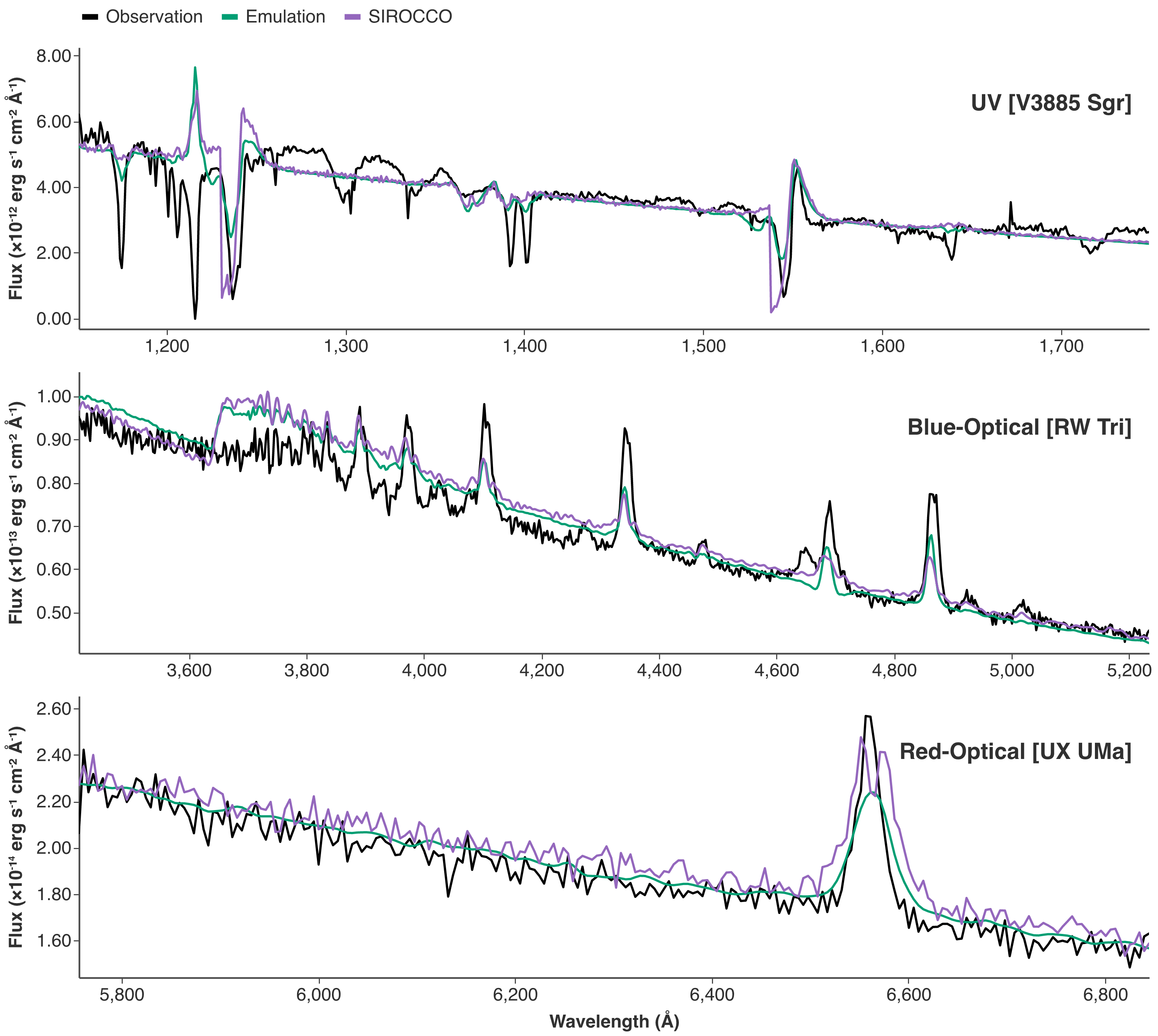}
    \caption{\textsc{speculate} discovered spectral models compared to observations using the Tier 3 Gaussian process emulator benchmark. The green line is the emulated spectrum from \textsc{speculate} at the MCMC median parameters, shown against the source observational spectrum in black. The purple line is the true \textsc{sirocco} spectrum generated from the emulated spectrum's parameters with the same nuisance parameter transformations applied, for example, the distance scaling, extinction corrections and Chebyshev polynomial continuum tilt. V3885 Sgr was fitted with our UV emulator configuration in Table~\ref{tab:emulator_config}, RW Tri with the Blue-Optical emulator, and UX UMa with the Red-Optical emulator. The fitting values can be found in Table~\ref{tab:obs_fit_values}.}
    \label{fig:obs}
\end{figure*}

For this paper, we are only going to focus on three of the five \textit{HST} spectra across wavelength ranges: V3885 Sgr in the \textbf{UV}, RW Tri (de-reddened) in the \textbf{Blue-Optical}, and UX UMa in the \textbf{Red-Optical}. This is partly for brevity, but it also allows us to explicitly demonstrate both the strengths and limitations of \textsc{speculate} and its underlying \textsc{sirocco} CV library. The emulated spectra produced by \textsc{speculate} at the MCMC median parameters, together with the matching \textsc{sirocco} spectra, are shown in Fig.~\ref{fig:obs}, with their corresponding marginalized MCMC median values in Table~\ref{tab:obs_fit_values} and posterior cornerplots in Figs.~\ref{fig:v3885sgr},~\ref{fig:rwtri}, and~\ref{fig:uxuma}. From these plots, we can make two deductions. Firstly, our emulated models appear accurate compared to the corresponding \textsc{sirocco} models, aside from the PCA approximation effects we discussed in section~\ref{sec:pca_truncation_artefacts}. Here, two of these can be seen clearly. (i) For the \ion{C}{iv} P-Cygni profile at 1550\,Å, the emulation has a modest slope across the blue absorption edge of the profile. Meanwhile, the real \textsc{sirocco} model has a much sharper, deeper edge. (ii) Several emulated blue emission lines tend to under- or overestimate the \textsc{sirocco} model. The emulated H\,$\alpha$ peak for UX UMa is also weaker than in the corresponding \textsc{sirocco} spectrum, although the two peaks were closer in strength when using the MCMC mean parameters.

These UV and blue-optical differences suggest that with the number of PCA components configured (30), we may be modelling too many complex spectral lines to achieve strong individual line recovery performance. The wavelength range may need to be reduced. Secondly, with \textsc{sirocco} models, we can fit multiple lines simultaneously to observations. However, some lines are better represented by our models than others, and some, like the \ion{Si}{ii}--\ion{O}{i} blended spectral line around 1300\,Å, are not modelled at all. To understand why this is, we need to consider the limitations of our underlying physical models: a real observational spectrum may not be purely disc-wind-dominated.

The \textsc{sirocco} CV grids released with \textsc{speculate} contain radiation from a blackbody-based accretion-disc SED, and optionally a blackbody boundary-layer SED, after it has propagated through and been reprocessed by the disc wind. The wind can therefore imprint both emission and absorption features on the underlying continuum. However, the adopted disc SED does not contain the intrinsic absorption features produced by a physical disc atmosphere, and the grids contain no spectral contribution from either the white dwarf or the donor star. Additionally, a specific aspect of this plot is that we used a non-boundary-layer grid, which does not include boundary-layer emission as a contribution to the extracted spectra. A real CV spectrum can include all of these components, together with contributions from hot spots in the accretion stream. At the same time, dust and the interstellar medium can further modify the observed spectrum. Therefore, we are not comparing like with like when fitting an observation that is not strongly disc-wind dominated. Wind-formed features are often prominent among the strong UV resonance lines, but at optical wavelengths the relative contribution of the wind can weaken.

Clear examples of this are Ly\,$\alpha$ at 1216\,Å and several blue-optical spectral lines in our other four demonstration \textit{HST} spectra. In these observations, the lines can appear as absorption lines. However, the present \textsc{speculate} CV grid can only produce these as emission lines. Broad absorption can be produced by the white-dwarf photosphere or the atmosphere of the optically thick disc, while Ly\,$\alpha$ may also contain a narrower absorption component from the interstellar medium. This is not a fundamental flaw in either \textsc{speculate} or \textsc{sirocco}, but is instead a consequence of our choice to use a blackbody-based disc SED that contains no intrinsic atmospheric absorption features. In essence, this example is physics-limited, not emulator-limited. \citet{matthews_impact_2015} previously showed that replacing the local blackbody spectrum of each disc annulus during spectral synthesis with a Kurucz stellar-atmosphere spectrum \citep{crivellari_new_1991} can introduce these absorption features. The fundamental restriction is therefore the input spectral library. With the \textsc{speculate} infrastructure already in place, nothing prevents other users or us from carefully constructing new \textsc{sirocco} grids that include these additional physical components.

Consequently, when the Tier 3 benchmark reports back the emulator/sirocco goodness-of-fit diagnostics and the posterior predictive checks (PPC), it is highly probable that good statistical scores will not be recovered in a traditional sense. For example, a reduced-$\chi^2$ near unity may not be recovered where the models cannot reproduce all observed features. Meanwhile, the PPC compares the observed spectrum with a 95 per cent band constructed from the MCMC model predictions. As we do not currently include observational noise or model uncertainty, the fraction of observed flux points need not be 95 per cent. Scores may improve with more heavily disc-wind-dominant sources. A researcher will likely find exploring the spectral fits alongside these diagnostics much more telling. Additionally, unconstrained MCMC corner plots may indicate that the fitted lines provide limited information about the wind parameters.

If we performed a fresh \textsc{speculate} iteration on these results with our current underlying grid, a few key changes could be made. To start, the wavelength ranges of the UV and Blue-Optical emulators could be reduced to exclude Ly\,$\alpha$ and the Balmer jump. Both of these spectral features are difficult for our grid to map to real observations. Therefore, we should not sacrifice emulator performance by including these lines when the freed-up power could instead be dedicated to other spectral lines. Additionally, for the red optical, the reverse may be true. We may be able to reduce the number of principal components configured to speed up training and inference. Next, for the recovered parameters, we may want to generate a new \textsc{sirocco} spectrum using the Kurucz stellar-atmosphere models \citep{crivellari_new_1991} instead of a blackbody SED. If the differences are significant, a researcher may want to create a new underlying CV grid focused on these models for observations that demand optical absorption, for example. 

One final technical point on fitting to observations: while (measured) errors on the spectral fluxes are recommended to be included, this is an optional requirement. If not present, \textsc{speculate} fits the continuum with a 2-degree polynomial in log-log space, with $3\,\sigma$ iterative clipping, so that spectral lines do not bias the continuum fit. An assumed error bar is then made, equal to 5 per cent of the continuum flux level ($\sigma(\lambda) = 0.05 \times F_{\mathrm{c}}(\lambda)$). Afterwards, \textsc{speculate} applies this corresponding continuum flux error to each wavelength measurement for its likelihood/covariance inference role\footnote{The keen reader may notice that \textsc{sirocco} test grid spectra are also available for inference workflows, and these too do not have errors. In this case, a 2 per cent continuum flux-level error is added to each measurement. 2 per cent is a rounded conservative estimate from the 95th-percentile robust fractional line-free continuum noise of all \textsc{sirocco} test grid spectra.}. 

\section{\textsc{speculate} user guide}
\label{sec:user guide}

\textsc{speculate} is distributed as an interactive \textsc{marimo} \citep{agrawal_marimo_2023} notebook suite with two deployment modes, online and locally, designed so that a user can browse the spectral library and obtain a fast fit in a web browser, or run the full training, inference, and benchmarking workflow on their own hardware. \textsc{marimo} is a `next-generation' Python notebook (similar to \textsc{jupyter notebooks}) that natively runs from Python files (.py) and is reactive: if a variable's value changes, other code dependent on that variable also updates (no hidden states). It allows notebooks to be deployed as apps with various built-in user interface features, such as sliders, drop-down menus, and drag-and-drop uploads. Data storage of the \textsc{sirocco} synthetic library is hosted on \textsc{hugging face} along with the online deployment of \textsc{speculate} as a \textsc{hugging face space}. Pre-trained emulators are also stored as \textsc{hugging face models}. Links to these resources and DOIs can be found in Data Availability. 

As of version 0.5.0, \textsc{speculate}'s functionality consists of six \textsc{marimo} notebooks: 
\begin{enumerate}
    \item \textbf{Model Downloader} -- This notebook is designed for downloading \textsc{sirocco} spectra and pre-trained models from \textsc{hugging face} for use in \textsc{speculate}. The data is stored in a compressed format on \textsc{hugging face}. The notebook automatically decompresses and places files in the correct location for other \textsc{speculate} notebooks to operate. This is guaranteed to be the first notebook a user will interact with when installing locally. 
    \item \textbf{Grid Inspector} \footnote{Only these two notebooks are available for use on the web. Also, a warning that the web-based version of \textsc{speculate} shares cached data between users, so uploaded data here will be visible to other users until the space falls asleep. If you have embargoed spectra, install \textsc{speculate} locally to fit protected data or for full functionality.\label{fn:web_note}} -- This notebook is designed for inspecting the \textsc{sirocco} spectral library. The user can browse through all the spectra and, with the interactive controls, transform the displayed spectrum to suit their individual needs. A user can also upload their own spectrum and manually search a grid space to find a closely matching simulated spectrum. When installed locally, enabling the \textsc{hugging face} mode switch in the sidebar allows for streaming of data through the \textsc{hugging face} API instead of directly downloading the entire library. However, reactivity is only as fast as the internet connection. 
    \item \textbf{Quick Fit} \textsuperscript{\ref{fn:web_note}} -- This notebook operates the full emulator training and inference workflow for \textsc{speculate}'s lightweight linear grid interpolation and neural network emulators, discussed in section~\ref{sec:Emu_GI} and~\ref{sec:Emu_NN}. When installed locally, a user can custom-configure and train an emulator to use in the inference stages, discussed in section~\ref{sec:chi-2 inference}. When online, the user is restricted to only the pre-trained emulator options stored on \textsc{hugging face}. A user can then test the performance of a trained emulator by running the inference ($\chi^2$ maximum likelihood estimation minimization) on the corresponding \textsc{sirocco} test grid spectra, but also add their own spectrum for this inference stage. Results of the best-fitting and associated parameters will be displayed and can be exported. Additionally, a \textsc{sirocco} parameter input file (.pf) can also be generated and used to run a true \textsc{sirocco} simulation of that particular parameter combination.
    \item \textbf{Training Tool} -- This notebook operates part 1 of 3 for \textsc{speculate}'s Gaussian Process emulator-based inference: the training of a custom configurable GP emulator, the details of which are discussed in section~\ref{sec:Emu_GP}. The notebook will analyse the users available computational resources and provide call-out warnings as necessary if a user is designing a too-heavyweight emulator for training and inference. Upon starting the training, the notebook will create the necessary grid and emulator .npz files, which are stored and shareable. Additionally, the notebook provides a live loss graph for monitoring training, terminal log information, and interactive Gaussian process diagnostics to inspect the fit to the PCA weight data. It displays the final emulator summary of GP variance and lengthscales. 
    \item \textbf{Benchmark Suite} -- This notebook operates part 2 of 3 of the GP emulator-based inference workflow: evaluating the performance of a trained emulator. This step is optional and not required to move from training to inference, but it allows the user to see if their emulator is performing adequately. Ultimately, a bad emulator will still work in the later inference stages, so without benchmarking, a user may be blissfully unaware of the bogus results it might output. As discussed in section~\ref{sec:Tier1}, a tier 1 benchmark should be the bare minimum that a user performs to evaluate a method. Tier 1 is a much faster benchmark than Tier 2 or Tier 3 (of the order of a minute or two) and should not be avoided. Tier 2 and Tier 3 benchmarks, as discussed in sections~\ref {sec:Tier2} and~\ref{sec:Tier3}, are also present within this notebook. Note: Tier 3 benchmarking requires a parallel installation of \textsc{sirocco} to run. All benchmarks include a suite of diagnostic plots and interactive inspection tools to evaluate performance; the details of these are documented on the \href{https://github.com/sirocco-rt/speculate/wiki}{GitHub Wiki}. All benchmark results are stored as JSON files so they can be reloaded later or shared.
    \item \textbf{Inference Tool} -- This notebook operates part 3 of 3 for the GP emulator-based inference workflow: fitting an emulated \textsc{sirocco} model to data. This tool allows for manual search of the parameter space through the parameter playground, or for optimizers to perform maximum-likelihood estimation and MCMC. Details of this process are discussed in section~\ref{sec:starfish inference}. Best-fitting results for the MLE and the MCMC posteriors are displayed, and results can be exported to CSV, along with a \textsc{sirocco} input parameter file (.pf) for verification. 
\end{enumerate}

Thanks to \textsc{marimo}'s app-like notebooks, it is extremely unlikely that a user will have to interact with any Python code to operate \textsc{speculate}. The remainder of \textsc{speculate}'s sidebar is dedicated to documentation links, where more in-depth information on how \textsc{speculate} operates is provided. 

\section{Discussion}
\label{sec:discussion}

The results above show that \textsc{speculate} has now reached the point of becoming a usable tool for fitting CV disc-wind spectra -- a major milestone for the \textsc{sirocco} code. Nevertheless, fitting spectra with a synthetic model remains challenging, and there are still many areas for improvement and clear directions for future work on the framework. These directions are as follows: 
\begin{enumerate}
    \item \textbf{Training data remains a dominant constraint} -- The quasi-regular $3^{N}$ grid (except inclination) is efficient and moderates the interpolation uncertainty, but is ultimately sparse in high dimensions. While these two grids cover a massive range of possible CV systems, the detailed physics modelled by our setup, such as the blackbody-based accretion-disc SED as discussed in \ref{sec:Tier3}, might not be suitable for every observation. However, from this setup, a researcher should be able to build more intuition about the \textsc{sirocco} modelling parameters and how they affect the resultant spectrum. Also, they can narrow down the region of parameter interest for a particular observation, or create their own smaller \textsc{sirocco} grid for this region and add it to the \textsc{speculate} framework to emulate. It might even be possible to skip this step entirely and move straight to modelling with \textsc{sirocco}. Along this front, \textsc{speculate}'s subsequent targets for a `better' CV grid would be problem-driven rather than general-purpose. In other words, designing grids to tackle specific problems. \textsc{sirocco} is still too computationally expensive to expand the training grid size much further beyond the limits imposed here, so adding additional physics is difficult both for \textsc{sirocco} and \textsc{speculate}'s GP emulator. A Latin Hypercube grid is a likely improvement over a regular grid for increasing the information density of a grid. However, some code structures in the methodology (regarding GP length scales and normalization-factor interpolations) will significantly increase computational time on an irregular grid. Hence, the improvements are not clear-cut. However, the neural network emulator version should not be affected by these points, so that it might provide a way forward. 
    \item \textbf{Gaussian Processes as a limiting factor} -- Ultimately, even with our performance improvements from the \textsc{starfish} framework, we are currently hard limited algorithmically and computational hardware-wise. The Gaussian processes in \textsc{speculate} are pushing the $\bigO(N^3)$ scaling in computational time and $\bigO(N^2)$ scaling in memory limits for all current-day hardware. This limit will improve with time as hardware improves.
    \item \textbf{More advanced neural network structures} -- \textsc{speculate} utilizes an MLP neural network emulator setup as a lightweight model. Hence, all researchers have the means to access fast \textsc{sirocco} models. However, the world of deep learning is massive, and the quick-fit notebook could be promoted to heavier-weight models or to a probabilistic role, with models like Bayesian neural networks serving as a finite approximation to a GP. This may be the most suitable investigative route to bypass GP bottlenecks in the near term. 
    \item \textbf{Improvements to the Log Likelihood function} -- The \textsc{starfish} log-likelihood function utilizing GP kernels to absorb model mismatches was a substantial reason \textsc{speculate} adopted this model. However, this flexibility has meant fitting disc-wind models specifically, where lines are highly variable, degenerate, and can translate across wavelength bins, which can be challenging. A specific example: sometimes the optimizer finds solutions in which blowing up the GP amplitude hyperparameter (the covariance matrix uncertainty) so that the model fit is uncertain everywhere, down-weighting the residuals, is preferred. This is even though the likelihood includes a penalty term for doing so! Exploring adaptations to this likelihood is a suitable area of investigation, such as introducing flux-based objectives, such as a Wasserstein distance. 
    \item \textbf{Calibration of the GP covariance matrix} -- We have seen in Table~\ref{tab:Tier2_coverage_probability} that it is common for our GP emulators to be (slightly) overconfident in their coverage probabilities. Root causes are still an area of active investigation. But a possible partial cause of this might be miscalibration of the GP's predictive covariance. One way to improve this is to estimate an appropriate scaling for the global kernel's covariance matrix to increase uncertainty via leave-one-out cross-validation. This will increase the emulator's total uncertainty contribution to the total error budget. However, this will only address uniform overconfidence and not differences in miscalibration between parameters or in the structure of shapes in per-parameter PP curves. While \citet{bachoc_cross_2013} finds LOO-CV can be effective for covariance-model misspecification, there are also particular weaknesses in highly regular grid designs, and calibration might not generalize well. So this point does not yet have a clear-cut solution.
    \item \textbf{Expansion of \textsc{speculate} to other accreting systems} -- This is the soonest approaching development. Closely followed by this paper, we are expanding \textsc{speculate} to an AGN grid of models. AGN in \textsc{sirocco} are much more computationally expensive than a CV model. As a result, \textsc{speculate} AGN emulators are a dramatic leap forward in unlocking the \textsc{sirocco} AGN parameter space for observational comparison. While saving the details for a later paper, the AGN training data is already available on the \textsc{hugging face} repository.
\end{enumerate}

While this list of `problems' can look extensive, it is worth noting the significance of the current developments:
\begin{itemize}
    \item We now have an effective route to fit \textsc{sirocco} models to observational data efficiently.
    \item An individual GP emulation takes a maximum of a few tens of milliseconds per output spectrum, up to a few milliseconds for a GI or NN emulator, all while maintaining good predictive accuracy and decent parameter recovery. This is a substantial speed-up over a typical simulation, meaning researchers can now effectively search \textsc{sirocco}'s parameter space, aiding understanding of how different physical wind parameters affect spectral line shapes and strengths.
    \item \textsc{speculate} is releasing a massive synthetic spectral library of CV accretion-disc wind spectra, open to all researchers.
    \item Finally, all this tooling is bundled into a highly interactive, easy-to-use interface, with a lightweight, no-install online version available, making reproducibility incredibly streamlined.
\end{itemize} 

\section{Summary}
\label{sec:summary}

In this paper, we have presented \textsc{speculate}, our emulator-based inference framework for \textsc{sirocco} accretion disc wind spectra, demonstrating its ability to fit observations with \textsc{sirocco} models promptly and effectively. \textsc{speculate} is bundled with a large library of synthetic CV disc-wind spectra, along with an easy-to-use Python user interface for easier community adoption. 

We began by motivating the need for fast synthetic disc wind models for observational comparison, leading to the introduction of emulation -- a machine learning model trained to approximate the output of a far more computationally expensive simulation -- which in \textsc{sirocco}'s case is typically on the order of a couple of CPU-core-hours per simulation. Emulators need training data, so this led to the \textsc{speculate} CV synthetic library. The library consists of $88\,680$ spectra ($\approx1$ million CPU-core-hours) across 8 varying physical (\textsc{sirocco}) parameters (plus inclination) -- disc accretion rate, wind mass-loss rate, a wind collimation parameter, wind mass-loss rate exponent, acceleration length scale, and exponent, boundary layer luminosity, and temperature -- split into two quasi-regular grids (datasets). One grid includes a boundary-layer emission contribution to the extracted spectrum, and the other does not. Both grids are publicly available in our \textsc{sirocco} organization \textsc{hugging face} repository (links in the Data Availability section).

We then introduced the \textsc{speculate} emulator architectures. This consisted of three flavours: a linear grid interpolation approach, a lightweight neural network approach, and a heavier-weight Gaussian process approach. Each approach interpolates the principal component analysis (PCA) weight space to emulate spectra. PCA is a lossy dimensionality reduction technique; hence, we showcased how these PCA approximations manifest themselves in a spectral grid. Afterwards, we showed how each emulator methodology is utilized in an inference framework. For the grid interpolation method (which trains near-instantly) and the neural network method (which can train on a laptop in an hour), predicting a new spectrum given a set of grid parameters takes a few milliseconds -- tying this sampling speed to a $\chi^2$ optimization yields best-fitting predictions within a couple of minutes on most devices. The Gaussian process is more computationally intensive in both training and prediction. However, the Gaussian process benefits from the ability to quantify its own emulator uncertainty and to design a more flexible likelihood function, originally developed in the \textsc{starfish} framework \citep{czekala_ianczestarfish_2018}. This flexible likelihood function means \textsc{speculate} can absorb systematic mismatches which are typical between a synthetic model and an observation. To implement this Gaussian process approach in a high-dimensional environment, we undertook significant algorithmic optimizations and GPU refactorisations to accelerate the original \textsc{starfish} code. The overall training and inference times on an A100 HPC GPU are around $40\,\mathrm{ms}$ per sample, taking on the order of an hour or two for a full inference pass. 

Given the vast number of possible emulator configurations that can be designed to meet a user's/observer's needs, we report example performance metrics for a few pre-trained Gaussian process emulator models, but primarily focus on the \textsc{speculate} benchmarking tools for user evaluation. The benchmark consists of three tiers: (i) How well does the emulator reproduce its own training data? (ii) How well does the emulator inference recover \textsc{sirocco} parameters from test grid spectra? (iii) How well can our emulator inference fit a \textsc{sirocco} model to an observational spectrum? Each tier consists of several diagnostic plots and statistics to report areas of accuracy and concern. At the time of this paper, we can re-create \textsc{sirocco} spectra with our emulators pretty well and recover physical parameters, albeit with some degree of overconfidence. The \textsc{sirocco} models can also fit some of our demonstration \textit{HST} spectra reasonably well, with caveats a user should be aware of. For example, modelling non-disc wind dominant lines, such as Ly\,$\alpha$, with a pure disc wind spectral line model will naturally lead to some inconsistencies. Readers should take care to understand which physics is included in these released grids and which is not. 

Finally, we provide a brief user guide to \textsc{speculate} and its \textsc{marimo} notebooks. \textsc{speculate} has a lightweight version (limited functionality) hosted online as a \textsc{hugging face space}, so that researchers can use pre-trained grid interpolation and neural network emulators for immediate observational inference (i.e., no installation required). While \textsc{speculate} still has some clear areas for further development regarding calibrating uncertainties, expanding training datasets and expanding the scope of accreting systems, \textsc{speculate} is a major step forward in designing a framework capable of fitting observational spectra with \textsc{sirocco} models.

\section*{Acknowledgements}

This work was supported by the Science and Technology Facilities Council [studentship project reference 2750006]. The authors acknowledge the use of the IRIDIS High Performance Computing Facility and associated support services at the University of Southampton in the completion of this work. James Matthews acknowledges funding from a Royal Society University Research Fellowship (URF$\backslash$R1$\backslash$221062). Partial support for Knox Long's effort on the project was provided by NASA through grant numbers HST-GO-16489 and HST-GO-16659 and from the Space Telescope Science Institute, which is operated by AURA, Inc., under NASA contract NAS 5-26555. Additional thanks are given to Professor Sebastian F. H\"{o}nig for use of his GPU computing resources during development of \textsc{speculate}. 

The authors acknowledge the use of Generative AI tools in the development of \textsc{speculate}, in particular: OpenAI's (\href{https://openai.com/codex/}{Codex/GPT 4.0--6.0}), Anthropic's (\href{https://claude.com/product/claude-code}{Claude Code/Opus 4.x series}), and Google's (\href{https://gemini.google.com/app?hl=en-GB}{Gemini/2.5 Pro--3.1 Pro}). These tools supported coding activities, asset image creation, and documentation of the code. We also used their feedback in writing this paper to flag formatting inconsistencies and passages possibly requiring correction. We retain responsibility for the research and the final content of this paper.

Authors declare no conflict of interest.

\section*{Data Availability}
\label{sec:Data Availability}

\textsc{sirocco}'s cataclysmic variable spectral library is hosted as \textsc{hugging face} data sets and can be found following our \href{https://huggingface.co/Sirocco-rt}{\textsc{sirocco} \textsc{hugging face} organizational page}. Data set DOIs are as follows: 

\begin{itemize}
    \item CV grid \textbf{including} boundary layer emission -- \href{https://huggingface.co/datasets/Sirocco-rt/speculate_cv_bl_grid_v87f?doi=true}{doi:10.57967/hf/7509},
    \item CV grid \textbf{without} boundary layer emission -- \href{https://huggingface.co/datasets/Sirocco-rt/speculate_cv_no-bl_grid_v87f?doi=true}{doi:10.57967/hf/7510},
    \item CV test grid \textbf{including} boundary layer emission -- \href{https://huggingface.co/datasets/Sirocco-rt/speculate_cv_bl_testgrid_v87f?doi=true}{doi:10.57967/hf/9720},
    \item CV test grid \textbf{without} boundary layer emission -- \href{https://huggingface.co/datasets/Sirocco-rt/speculate_cv_no-bl_testgrid_v87f?doi=true}{doi:10.57967/hf/7512}.
\end{itemize}

A set of pre-trained CV emulators is available immediately for inference and stored in a \href{https://huggingface.co/Sirocco-rt/speculate_cv_emulators}{\textsc{hugging face} model repository}. Datasets and models can be easily downloaded through \textsc{speculate}'s `Model Downloader' tool. A tool restricted version of \textsc{speculate} (lightweight inference and dataset browser) can be found online, hosted as a \href{https://huggingface.co/spaces/Sirocco-rt/Speculate}{\textsc{hugging face} space}. If the space is asleep due to inactivity, click `restart the space` to rebuild the Docker container (usually takes a couple of minutes). Alternatively, the full feature version of \textsc{speculate} can be installed by following \citet{wallis_speculate_2026}: \textsc{speculate}'s \href{https://github.com/sirocco-rt/speculate}{\textsc{github} repository}. \textsc{speculate}'s documentation is hosted as a \href{https://github.com/sirocco-rt/speculate/wiki}{\textsc{github} Wiki}. Plots generated in this research work used several versions of \textsc{speculate}, up to v0.5.0. 

\textsc{sirocco} is a freely available software via \href{https://github.com/sirocco-rt/sirocco}{\textsc{github}} with documentation on \href{https://sirocco-rt.readthedocs.io/en/}{ReadTheDocs}. These CV synthetic library data sets utilise a pre-release version of \textsc{sirocco}, formally known as \textsc{Python v87f}. The original \textsc{starfish} framework code can be found following \citet{czekala_ianczestarfish_2018}.

Any identified bugs can be reported through the corresponding software's \textsc{github} repository.
\\



\bibliographystyle{mnras}
\bibliography{references} 


\appendix

\section{The 5 pre-loaded demo \textit{HST} observational spectra distances and inclinations}

Parsec distances are determined from \textit{Gaia} parallax measurements using a simple inverse-parallax approximation method following the equations:
\begin{equation}
    d\,[\mathrm{pc}] = 1\,000/p\,[\mathrm{mas}] ,
    \qquad
    \sigma_d = 1\,000\,\sigma_p/p^2
\end{equation}
where $d$ is distance in parsecs, $p$ is parallax in milliarcseconds, and $\sigma$ the error of the corresponding parameter. We use only a simple approximation here rather than more sophisticated methods, such as \citet{bailer-jones_estimating_2021}, as \textsc{speculate} will not be sensitive enough to warrant such accurate, asymmetric errors. Inclinations are sourced from literature. 

\begin{table*}
\centering
\small
\setlength{\tabcolsep}{4pt}

    \begin{tabular}{cccccc}
\hline
        Source & \textit{Gaia} DR3 ID & \makecell{\textit{Gaia} parallax\\(mas)} & \makecell{Approximate distance\\($\mathrm{pc}$)} & \makecell{Inclination\\($^{\circ}$)} & \makecell{Inclination\\citations} \\
        \hline
        IX Vel & 5515820034889610112 & $11.0674 \pm 0.0182$ & $90.36 \pm 0.15$ & $57 \pm 2$ & \citep{linnell_synthetic_2007}\\
        RW Sex & 3769067109159364608 & $4.4819 \pm 0.0246$ & $223.12 \pm 1.23$ & 28--40  & \citep{beuermann_phase-resolved_1992} \\
        RW Tri & 130692247044752768 & $3.2671 \pm 0.0223$ & $306.08 \pm 2.09$ & $72.5 \pm 2.5$ & \citep{smak_rw_2019} \\
        UX UMa & 1559987685901122304 & $3.4252 \pm 0.0157$ & $291.95 \pm 1.34$ & $71.0 \pm 0.6$ & \citep{baptista_hubble_1995} \\
        V3885 Sgr & 6688624794231054976 & $7.7535 \pm 0.0344$ & $128.97 \pm 0.57$  & $45\text{--}70$ & \citep{ribeiro_emission-line_2007}\\
\hline
\end{tabular}
\caption{A compiled list of parsec distances and inclinations used for \textsc{speculate}'s 5 demo \textit{HST} observations of nova-like variables. Parallaxes are retrieved from \textit{Gaia} Early Data Release 3 \citep{gaia_collaboration_gaia_2021}.}
\label{tab:distances_and_inclinations}
\end{table*}

\section{Tier 1 -- Training-grid RMSE envelopes}
\label{sec:Tier 1-- Training grid RMSE envelopes}

\begin{figure*}
\centering

	\includegraphics[width=\linewidth,height=0.74\textheight,keepaspectratio,
        alt={Three panels plot training-grid reconstruction RMSE against wavelength in the ultraviolet, blue-optical and red-optical ranges. The PCA truncation curve is compared with grid interpolation, neural-network and Gaussian-process curves computed using their different evaluation methods.}
    ]{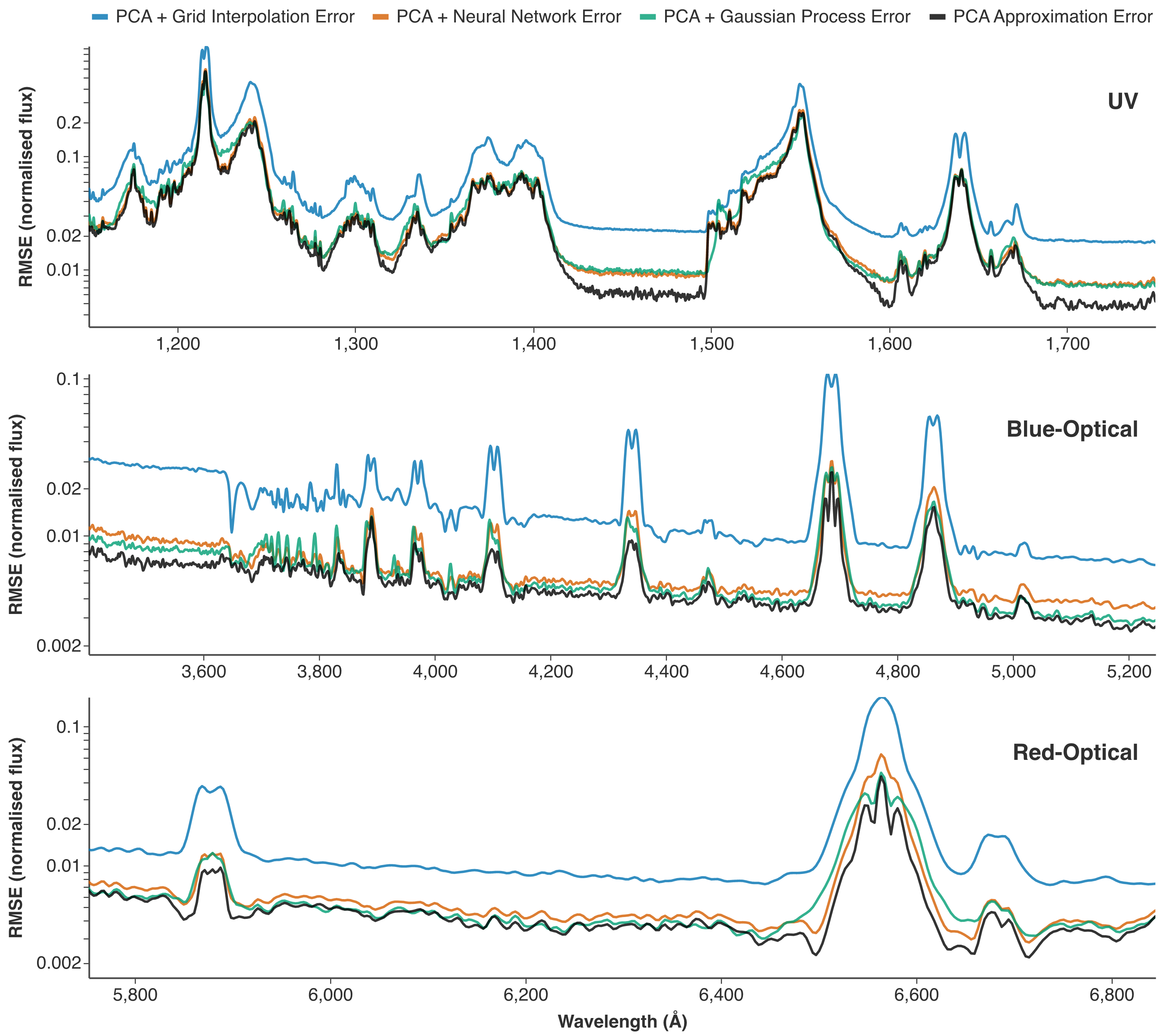}
    \caption{The per-wavelength training-grid flux RMSE for the UV, blue-optical, and red-optical emulator configurations given in Table~\ref{tab:emulator_config}. The black line shows the reconstruction error obtained from a PCA truncation at 30 principal components. The blue, orange, and green lines show the total reconstruction error caused by the grid interpolation (GI), the neural network (NN), and the Gaussian Process (GP) method, respectively. The GI curve is calculated using an approximate leave-one-out (LOO) methodology, the NN using an in-sample methodology, and the GP using the LOO methodology. All spectra are mean-normalized before RMSE calculations and, due to the different evaluation methodologies, should not be used for comparison between architectures and are not indicative of emulator performance off the training grid points.}
    \label{fig:train-RMSE-Tier1}
\end{figure*}

In section~\ref{sec:Tier1}, we discuss the main method a user can use to see whether an emulator has weak performance on its own data. This was through plotting the per-wavelength RMSE (root mean square error) envelope. Here, we present the training-grid version of this plot. This plot was placed in the appendix as there is no clean approach across architectures to provide a like-for-like comparison. Each architecture had to use slightly different evaluation methods, so RMSE curves in Fig.~\ref{fig:train-RMSE-Tier1} should not be used to infer indicative performance between architectures. 

For the Gaussian process RMSE curve, we use the leave-one-out (LOO) methodology. For a given grid point under evaluation, LOO removes that point and then attempts to predict it using the remaining points. Sounds computationally expensive! However, there is a trick to avoid retraining entirely. The prediction with that point omitted is calculated via an inverse-covariance identity, keeping the fitted GP hyperparameters and PCA preprocessing fixed. Performing this procedure for all data points and comparing the resulting reconstructed spectra against their true training-grid spectra gives a final LOO-CV per-wavelength RMSE (see sections 5.3 and 5.4.2 in \citet{rasmussen_gaussian_2006}). This tests predictions with each point omitted from the GP conditioning, rather than refitting the full emulator for every omitted point.

Unfortunately, for the other architectures, we do not have this trick, and an LOO evaluation would require retraining at each grid point, which is computationally prohibitive. So instead, the grid interpolation method utilizes an approximate LOO methodology, in which the removed grid point is estimated as the mean of its immediately adjacent grid points along each parameter axis. The neural network architecture differs in that it is not constrained to pass through the grid points in the PCA weight space. Hence, for the NN, we can emulate it using an emulator trained on the full grid and evaluate it at the grid points. This leaves an `in-sample' residual, i.e., how well does the NN represent the input grid? These differences mean we cannot explicitly compare against different architectural methods, only with the same architectures across different emulator configurations. In other words, one grid interpolation emulator against another grid interpolation emulator.  

A plot of these measures can be seen for our UV, Blue-Optical and Red-Optical configurations in Fig.~\ref{fig:train-RMSE-Tier1}. This plot shows the reconstruction error for each emulator method at every wavelength bin, with the PCA truncation error shown in black.

\clearpage
\begin{table*}
\noindent\begin{minipage}{\dimexpr(\textwidth-\columnsep)/2\relax}
\normalsize
\section{Tier 3 -- MCMC median values}
\label{sec:Tier 3 -- Best Fit MCMC Values}

Table~\ref{tab:obs_fit_values} lists the median MCMC physical parameters and their uncertainties for the three observational fits shown in Fig.~\ref{fig:obs}. The corresponding posterior corner plots are shown in Figs.~\ref{fig:v3885sgr},~\ref{fig:rwtri}, and~\ref{fig:uxuma}.
\par
\end{minipage}\hfill\mbox{}\par
\vspace{2\baselineskip}
\centering
\renewcommand{\arraystretch}{1.4}
    \begin{tabular}{lllll}
\hline
    
         \textbf{Physical Parameter} & \textbf{Unit} & \textbf{UV [V3885 Sgr]} & \textbf{Blue-Optical [RW Tri]} & \textbf{Red-Optical [UX UMa]} \\
         \hline 
         \textbf{Disc Accretion Rate} &$M_{\odot}\,\mathrm{yr}^{-1}$ & $7.05_{-1.19}^{+1.97}\times10^{-9}$  & $6.37_{-1.11}^{+1.00}\times10^{-9}$ & $2.58_{-0.65}^{+0.30}\times10^{-8}$ \\
         \textbf{Wind Mass Loss Rate} &$M_{\odot}\,\mathrm{yr}^{-1}$ & $8.13_{-4.44}^{+6.38}\times10^{-10}$ & $1.33_{-0.37}^{+0.36}\times10^{-9}$ & $1.03_{-0.19}^{+0.44}\times10^{-9}$ \\
         \textbf{Wind Collimation}& $R_{\mathrm{WD}}$ & $23.81_{-9.77}^{+8.55}$ & $1.12_{-0.38}^{+1.08}$ & $5.70_{-2.55}^{+7.05}$ \\
         \textbf{Mass Loss Rate Exponent}& -- & $0.15_{-0.11}^{+0.22}$ & $0.21_{-0.16}^{+0.33}$ & $0.70_{-0.20}^{+0.14}$ \\
         \textbf{Acceleration Length Scale}& $\mathrm{cm}$ & $9.31_{-4.94}^{+24.02}\times10^{9}$ & $2.62_{-1.09}^{+1.85}\times10^{10}$ & $5.19_{-3.80}^{+1.41}\times10^{10}$\\
         \textbf{Acceleration Exponent} &-- & $3.11_{-1.43}^{+0.93}$ & $3.60_{-0.81}^{+0.63}$ & $1.68_{-0.29}^{+0.75}$ \\
         \textbf{Inclination}& ° & $48.35_{-2.33}^{+4.16}$ & $69.71_{-1.80}^{+1.45}$ & $71.03_{-0.53}^{+0.53}$ \\
\hline
\end{tabular}
\caption{The MCMC median values corresponding to the emulated spectra in Fig.~\ref{fig:obs}. As a reminder, these values should not be used as publishable values for these systems. These values should only be used as starting points for further modelling with \textsc{sirocco}. Values are rounded to 2 decimal places.}
\label{tab:obs_fit_values}
\end{table*}

\begin{figure*}
\centering
    \includegraphics[width=\linewidth,height=0.85\textheight,keepaspectratio,
        alt={A corner plot that shows the one- and two-dimensional marginal posterior distributions of seven wind parameters for the ultraviolet fit to V3885 Sgr. Several parameters have broad and/or asymmetric distributions with some hints of multi-modalities present.}
    ]{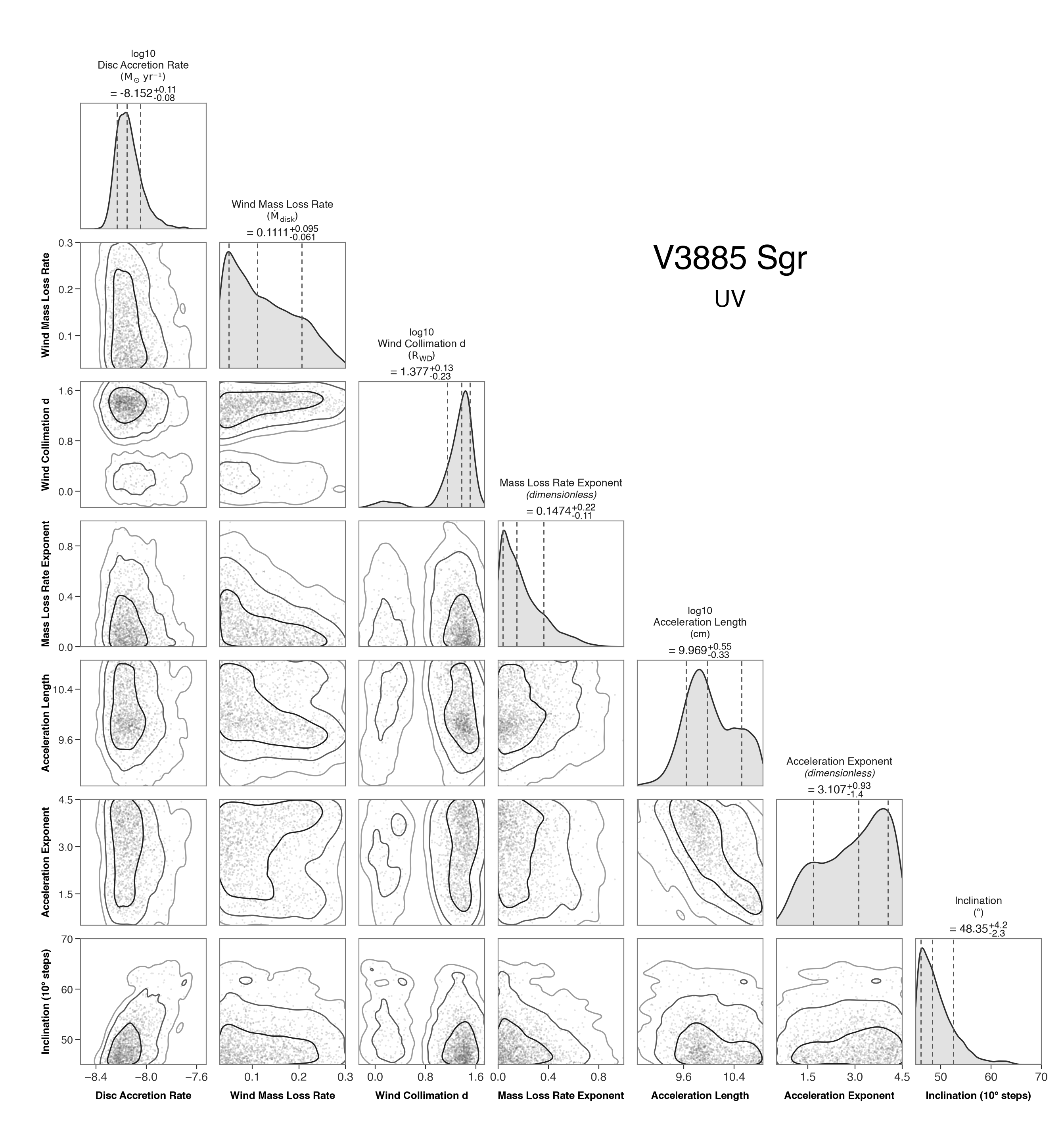}
    \caption{An MCMC corner plot showing the marginalized 1D and 2D posterior probability distributions for V3885 Sgr's \textsc{sirocco} model parameters. The 2-D plot contours enclose 68.27 per cent, 95.45 per cent, and 99.73 per cent of each pairwise marginal posterior. The vertical black dashed lines highlight the median and the $1\,\sigma$ credible interval of each diagonal 1-D probability density function. The axes show the adopted prior ranges, with Gaussian inclination priors displayed over $\pm4\,\sigma$ about their prior means. Strong correlations between parameters can indicate degeneracies. Separate peaks may indicate multimodality, with multiple possible solutions to a single fit.}
    \label{fig:v3885sgr}
\end{figure*}

\begin{figure*}
\centering
    \includegraphics[width=\linewidth,height=0.85\textheight,keepaspectratio,
        alt={A corner plot that shows the one- and two-dimensional marginal posterior distributions of seven wind parameters for the blue-optical fit to RW Tri. Several parameters have broad and/or asymmetric distributions with some hints of multi-modalities.}
    ]{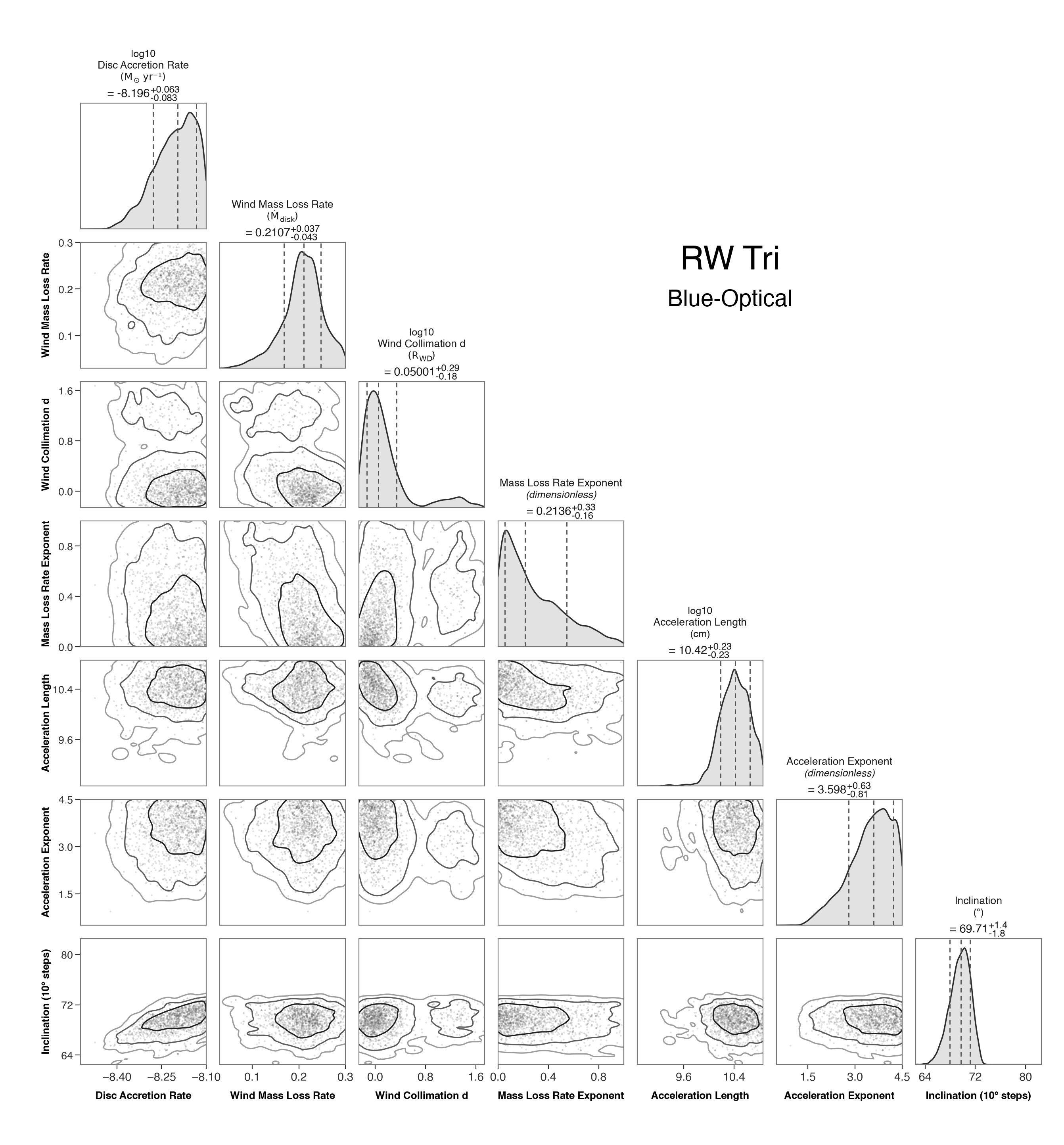}
    \caption{Same as Fig.~\ref{fig:v3885sgr} but for RW Tri.}
    \label{fig:rwtri}
\end{figure*}

\begin{figure*}
\centering
    \includegraphics[width=\linewidth,height=0.85\textheight,keepaspectratio,
        alt={A corner plot that shows the one- and two-dimensional marginal posterior distributions of seven wind parameters for the red-optical fit to UX UMa. Several parameters have broad and/or asymmetric distributions.}
    ]{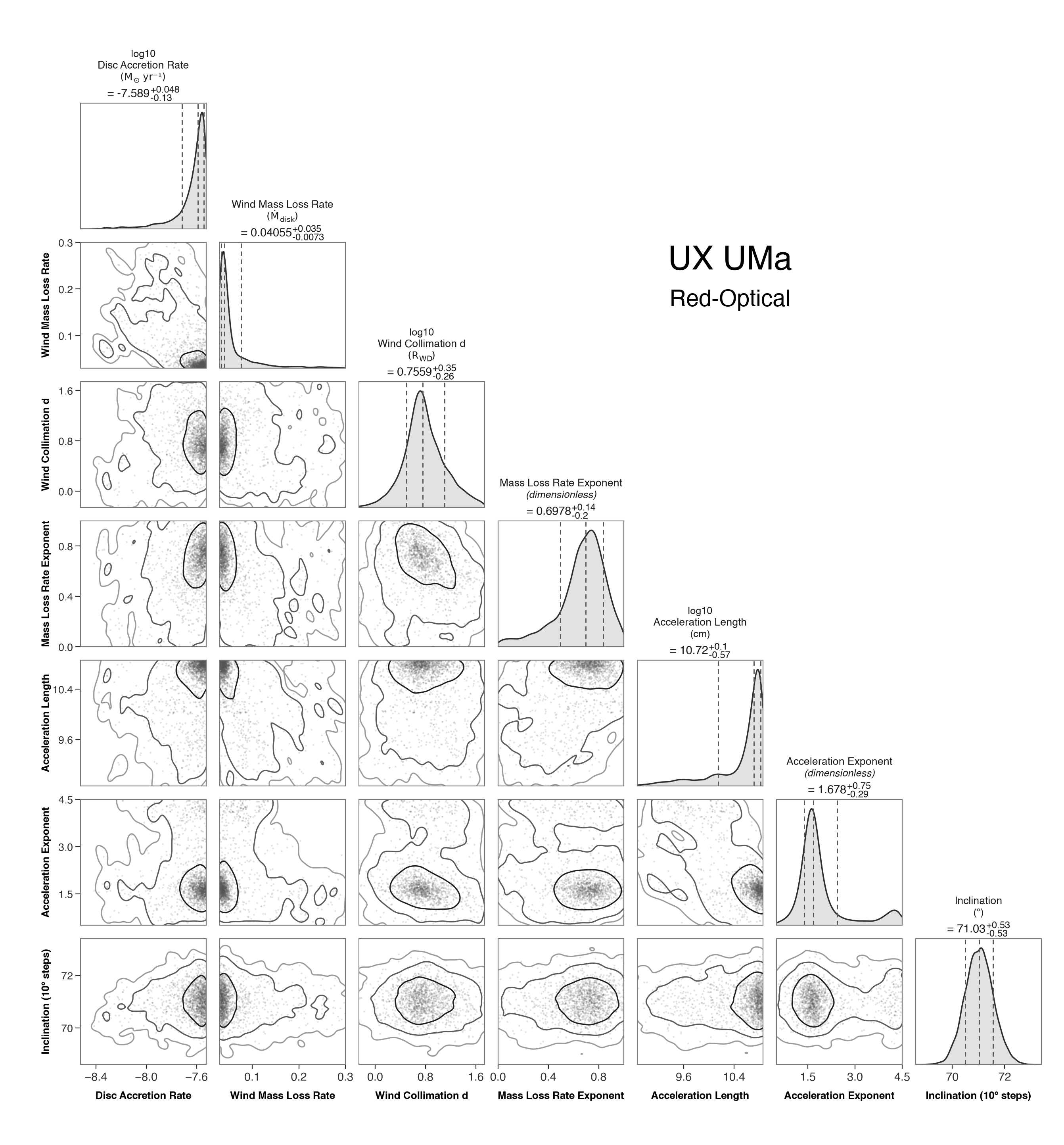}
    \caption{Same as Fig.~\ref{fig:v3885sgr} but for UX UMa.}
    \label{fig:uxuma}

{\raggedright\bsp\par}
\label{lastpage}
\end{figure*}

\end{document}